\documentclass[12pt,eqsecnum]{article}
\usepackage[dvips]{graphicx}
\usepackage{amssymb}
\usepackage{amsmath}
\usepackage{ulem}
\usepackage{epstopdf}
\usepackage[dvipsnames]{xcolor}
\usepackage{soul}
\usepackage{cite}
\usepackage{mathtools}

\makeatletter

\@addtoreset{equation}{section}
\makeatletter

\newcommand{\ma}[1]{\mbox{$\mathcal{#1}$}}
\newcommand{\qed}{\hbox{\rule[-2pt]{6pt}{6pt}}}
\newcommand{\D}{{\rm d}}

\newtheorem{Prop}{Proposition}

\newtheorem{lm}{Lemma}
\newtheorem{dn}{Definition}

\newcommand{\dalm}{\kern1pt\vbox{\hrule height 0.9pt\hbox{\vrule width
0.9pt\hskip 2.5pt\vbox{\vskip 5.5pt}\hskip 3pt\vrule width 0.3pt}\hrule height
0.3pt}\kern1pt}

\begin{document}

\begin{titlepage}
\vfill
\begin{flushright}
\today
\end{flushright}

\vfill
\begin{center}
\baselineskip=12pt
{\Large\bf 
Existence and absence of Killing horizons in static solutions with symmetries
}
\vskip 0.5cm
{\large {\sl }}
\vskip 5.mm
{\bf Hideki Maeda${}^{a,b}$\footnote{Author to whom any correspondence should be addressed.} and Cristi{\'a}n Mart\'{\i}nez$^{c,d}$} \\

\vskip 1cm
{
${}^a$Department of Electronics and Information Engineering, Hokkai-Gakuen University, Sapporo 062-8605, Japan.\\
${}^b$Max-Planck-Institut f\"ur Gravitationsphysik (Albert-Einstein-Institut), \\Am M\"uhlenberg~1, D-14476 Potsdam, Germany.\\
${}^c$Centro de Estudios Cient\'{\i}ficos (CECs), Av. Arturo Prat 514, Valdivia, Chile. \\
${}^d$Facultad de Ingenier\'{\i}a, Arquitectura y Dise\~no, Universidad San Sebasti\'an, \\General Lagos 1163, Valdivia, 5110693, Chile.\\
\texttt{h-maeda@hgu.jp, cristian.martinez@uss.cl}
}
\vspace{6pt}
\end{center}
\vskip 0.2in
\par
\begin{center}
{\bf Abstract}
\end{center}
\begin{quote}
Without specifying a matter field nor imposing energy conditions, we study Killing horizons in $n(\ge 3)$-dimensional static solutions in general relativity with an $(n-2)$-dimensional Einstein base manifold.
Assuming linear relations $p_{\rm r}\simeq\chi_{\rm r} \rho$ and $p_2\simeq\chi_{\rm t} \rho$ near a Killing horizon between the energy density $\rho$, radial pressure $p_{\rm r}$, and tangential pressure $p_2$ of the matter field, we prove that any non-vacuum solution satisfying $\chi_{\rm r}<-1/3$ ($\chi_{\rm r}\ne -1$) or $\chi_{\rm r}>0$ does not admit a horizon as it becomes a curvature singularity. 
For $\chi_{\rm r}=-1$ and $\chi_{\rm r}\in[-1/3,0)$, non-vacuum solutions admit Killing horizons, on which there exists a matter field only for $\chi_{\rm r}=-1$ and $-1/3$, which are of the Hawking-Ellis type~I and type~II, respectively.
Differentiability of the metric on the horizon depends on the value of $\chi_{\rm r}$, and non-analytic extensions beyond the horizon are allowed for $\chi_{\rm r}\in[-1/3,0)$.
In particular, solutions can be attached to the Schwarzschild-Tangherlini-type vacuum solution at the Killing horizon in at least a $C^{1,1}$ regular manner without a lightlike thin shell. 
We generalize some of those results in Lovelock gravity with a maximally symmetric base manifold.
\vskip 1.mm
\noindent
{\bf Note added:} The present arXiv version has Appendix~\ref{addendum} based on the addendum (2025 Class. Quantum Grav. {\bf 42}, 129401) to the published version (2024 Class. Quantum Grav. 41, 245013).
\vfill
\vskip 2.mm
\end{quote}
\end{titlepage}



\tableofcontents

\newpage

\section{Introduction}

While black holes are defined by the existence of an event horizon, the rigidity theorem asserts that the event horizon of a stationary black hole is a Killing horizon~\cite{arealaw,Hawking:1973uf,Chrusciel:1996bj,Hollands:2006rj}.
The analysis of the Killing horizon then led to the formulation of the black hole thermodynamics for stationary black holes~\cite{Bardeen1973,Hawking:1974sw,wald2,Wald:1993nt,iyerwald1994,Iyer:1995kg}.
Those results are valid not only in four dimensions but also in higher dimensions.
However, the regularity of the Killing horizon may vary with the number of spacetime dimensions.

The higher dimensional counterparts of the Schwarzschild and Kerr vacuum black holes, the Tangherlini and Myers-Perry black holes, have Killing horizons.
In contrast, regularity of the Killing horizon in static solutions~\cite{Majumdar:1947eu,Papaetrou:1947ib,Lemos:2005md} describing multiple extremal black holes in the Einstein-Maxwell system sharply depends on the number of dimensions $n$.
While the metric of the Majumdar-Papapetrou solution for $n=4$ is $C^\infty$ at the Killing horizon~\cite{Hartle:1972ya}, the metric of its higher-dimensional counterpart is $C^2$ at the horizon for $n=5$, and the Killing horizon turns into a parallelly propagated (p.p.) curvature singularity for $n\ge 6$~\cite{Candlish:2007fh}. (See also Ref.~\cite{Welch:1995dh}.)
Recently, such a transition of the Killing horizon to a singularity in higher dimensions has been observed in a static perfect-fluid solution obeying a linear equation of state $p=-(n-3)\rho/(n+1)$~\cite{Maeda:2022lsm}.
The Killing horizon in that solution is regular with a $C^\infty$ metric for $n=4$ and $5$, but turns into a p.p. curvature singularity for $n\ge 6$.

The question of when a Killing horizon becomes regular and extendible has been studied without specifying a matter field~\cite{Pravda:2005uv,Zaslavskii:2007pg,Bronnikov:2008ia,Bronnikov:2008by,Bronnikov:2009ui,Bronnikov:2011nb}.
Such a black hole in interaction with matter fields is sometimes called a {\it dirty black hole}~\cite{Visser:1992qh}.
As shown by Horowitz and Ross earlier~\cite{Horowitz:1997uc,Horowitz:1997ed}, the tidal force in a free-falling frame towards the horizon of such a dirty black hole can be enormous even if the curvature polynomials remain finite there, so that the quantum effects of gravity could be significant near a horizon even for a large black hole.
Such a black hole is referred to as a {\it naked black hole} and in fact realized in various systems in general relativity~\cite{Horowitz:1997uc,Horowitz:1997ed}.

In 2005, Pravda and Zaslavskii studied a general four-dimensional static spacetime and referred to a spacetime with a Killing horizon where the Kretschmann scalar is finite but the Weyl scalars $\Psi_3$ and/or $\Psi_4$ diverge as a ``truly naked black hole''~\cite{Pravda:2005uv}. However, the truly naked horizon is actually a p.p. curvature singularity and therefore inextendible as the authors of that paper clearly stated in the introduction.
The same terminology was also used  in Refs.~\cite{Zaslavskii:2007pg,Bronnikov:2008by}.
In Refs.~\cite{Bronnikov:2008ia,Bronnikov:2008by,Bronnikov:2009ui,Bronnikov:2011nb}, it was investigated by asymptotic analyses when a Killing horizon becomes regular and extendible in four dimensions. In those references, the regularity of the horizon is defined by analyticity ($C^\omega$) of the metric, although the authors briefly mentioned the possibility of extension with lower differentiability of the metric\footnote{Indeed, $C^\omega$ (the Taylor series is convergent) is too strong in general relativity because spacetime is regular and extendible if the metric is $C^{1,1}$ (often denoted by $C^{2-}$ in physics), which is sufficient to compute curvature tensors.}. 
Specifically in Ref.~\cite{Bronnikov:2008ia}, under the assumptions near a Killing horizon (i) analyticity of the metric ($C^\omega$), (ii) the null energy condition, and (iii) $p_{\rm r}\simeq\chi_{\rm r} \rho$ and $|p_2/\rho|\ll \infty$ for the energy density $\rho$, radial pressure $p_{\rm r}$, and tangential pressure $p_2$ of the matter field, the authors showed that there exist static solutions for $\chi_{\rm r}=-1/(1+2N)$ with a natural number $N$.

The present article follows on from the above studies and aims to provide new results cleanly.
We generalize those studies for $n(\ge 3)$-dimensional static solutions in general relativity with an $(n-2)$-dimensional Einstein base manifold, without specifying a matter field nor imposing energy conditions. 
The existence or absence of a Killing horizon is established only under the assumption of a linear relation between the pressures and the density near the horizon.
In some propositions, we consider assumptions that are different from the previous studies to make the results more mathematically rigorous. 
We also extend some of the main results to Lovelock gravity~\cite{Lovelock:1971yv}, which is the most natural generalization of general relativity in arbitrary dimensions without torsion such that the field equations are of the second order.

The organization of the present article is as follows.
In Sec.~\ref{sec:Preliminaries}, we review several mathematical concepts and prove useful lemmas for the subsequent section. 
Our main results are presented in Sec.~\ref{sec:GR}. 
We summarize our main results and give concluding remarks in the final section.
In Appendix~\ref{app:comple}, we provide several corrections and complements to the results in Sec.~IV in Ref.~\cite{Bronnikov:2008by}.
In Appendix~\ref{app:exact}, we present some exact solutions which complement the results in Sec.~\ref{sec:GR}.
Appendix~\ref{app:coefficients} details the proof of Proposition~\ref{Prop:asymp}.
Our conventions for curvature tensors are $[\nabla _\rho ,\nabla_\sigma]V^\mu ={{R}^\mu }_{\nu\rho\sigma}V^\nu$ and ${R}_{\mu \nu }={{R}^\rho }_{\mu \rho \nu }$, where Greek indices run over all spacetime indices.
The signature of the Minkowski spacetime is $(-,+,+,\cdots,+)$ and other types of indices will be specified in the main text.
The sets of natural numbers are denoted by $\mathbb{N}(=1,2,3,\cdots)$ and $\mathbb{N}_0(=0,1,2,\cdots)$.
We adopt the units such that $c=1$ and $\kappa_n:=8\pi G_n$, where $G_n$ is the $n$-dimensional gravitational constant.
Throughout this paper, a prime denotes differentiation with respect to a coordinate $x$.

\section{Static spacetime with symmetries}
\label{sec:Preliminaries}

In the present article, we study $n(\ge 3)$-dimensional static spacetimes with an $(n-2)$-dimensional Einstein base-manifold $K^{n-2}$.
For this purpose, we use the metric 
\begin{align}
&\D s^2=-H(x)\D t^2+\frac{\D x^2}{H(x)}+r(x)^2\gamma_{ij}(z)\D z^i\D z^j\label{metric-Buchdahl}
\end{align} 
in the quasi-global coordinates $x^\mu=(t,x,z^i)$~\cite{Bronnikov:2008ia} and we assume $r(x)\ge 0$ without loss of generality, where $z^i$ and $\gamma_{ij}$ ($i,j=2,3,\cdots,n-1$) are coordinates and a metric on $K^{n-2}$, respectively.
A region with $H>(<)0$ is static (dynamical) since a hypersurface-orthogonal Killing vector $(\partial/\partial t)^\mu$ is timelike (spacelike).

The Riemann tensor of the Einstein space $K^{n-2}$ can be written as
\begin{align}
{}^{(n-2)}{R}_{ijkl}={}^{(n-2)}{C}_{ijkl}+k(\gamma_{ik}\gamma_{jl}-\gamma_{il}\gamma_{jk}),\label{Weyl}
\end{align} 
where $k$ is a constant and ${}^{(n-2)}{C}_{ijkl}$ is the Weyl tensor on $K^{n-2}$~\cite{Dotti:2005rc}.
The Ricci tensor of $K^{n-2}$ is given by ${}^{(n-2)}R_{ij}=k(n-3)\gamma_{ij}$ and $K^{n-2}$ is maximally symmetric if and only if ${}^{(n-2)}{C}_{ijkl}\equiv 0$.
By redefining the areal radius $r(x)$, the constant $k$ can be set to take $1$, $0$, and $-1$ without loss of generality corresponding to positive, zero, and negative curvature of $K^{n-2}$, respectively.
For $n=3$, we have $k=0$ and ${}^{(1)}{C}_{ijkl}\equiv 0$.
For $n=4$ and $5$, we have ${}^{(n-2)}{C}_{ijkl}\equiv 0$, so that $K^{n-2}$ is maximally symmetric. 
In this section, we summarize basic properties of static spacetimes described by the metric (\ref{metric-Buchdahl}) and the corresponding energy-momentum tensor.
It is emphasized that most of the following arguments are theory-independent.

In Ref.~\cite{Buchdahl1954}, Buchdahl introduced a coordinate system for the most general static spacetime in $n(\ge 4)$ dimensions in which the metric is written as 
\begin{align}
\label{gauge-higher}
\D s^2=-\Omega(X)^{-2}\D t^2+\Omega(X)^{2/(n-3)}{\bar g}_{IJ}(X)\D X^I\D X^J,
\end{align}
where $I,J=1,2,\cdots,n-1$.
Hence, the quasi-global coordinates (\ref{metric-Buchdahl}) and the Buchdahl coordinates (\ref{metric-Buchdahl}) coincide in four dimensions~\cite{Boonserm:2007zm}.

\subsection{Killing horizons and null infinities}

In the coordinate system (\ref{metric-Buchdahl}), there is a coordinate singularity $x=x_{\rm h}$ determined by $H(x_{\rm h})=0$.
It is avoided by introducing an ingoing null coordinate $v:=t+\int H(x)^{-1}\D x$ and then the metric (\ref{metric-Buchdahl}) turns into\footnote{Alternatively, one may use an outgoing-null coordinate $u:=t-\int H(x)^{-1}\D x$ and then the resulting metric is given by Eq.~(\ref{metric-Buchdahl-v}) with $v$ replaced by $-u$.}
\begin{equation}
\D s^2=-H(x)\D v^2+2\D v\D x+r(x)^2\gamma_{ij}(z)\D z^i\D z^j,\label{metric-Buchdahl-v}
\end{equation}
of which inverse metric is given by
\begin{align}
\begin{aligned}
&g^{vv}=0,\qquad g^{vx}=g^{xv}=1,\\
&g^{xx}=H,\qquad g^{ij}=r^{-2}\gamma^{ij}.
\end{aligned}
\end{align}
In the single null coordinates (\ref{metric-Buchdahl-v}), $x=x_{\rm h}$, where a hypersurface-orthogonal Killing vector $\xi^\mu=(\partial/\partial v)^\mu$ becomes null, is a Killing horizon if it is regular.
\begin{dn}[Killing horizon]
\label{dn:K-horizon}
In a spacetime described by the metric (\ref{metric-Buchdahl-v}), a {\it Killing horizon} associated with a Killing vector $\xi^\mu=(\partial/\partial v)^\mu$ is a regular null hypersurface $x=x_{\rm h}$, where $x_{\rm h}$ is determined by $H(x_{\rm h})=0$ with $r(x_{\rm h})\ne 0$.
A Killing horizon is referred to as non-degenerate (degenerate) if $H'(x_{\rm h})\ne 0$ ($H'(x_{\rm h})= 0$) holds.
\end{dn}

It should be emphasized that, as demonstrated in Ref.~\cite{Maeda:2021ukk}, singular coordinates on the horizon, such as the quasi-global coordinates (\ref{metric-Buchdahl}), do not allow us to calculate physical and geometrical quantities on the horizon properly.
The Gaussian normal coordinates are also such coordinates given by 
\begin{equation}
\D s^{2}=-N^{2}\D t^{2}+\D n^{2}+{\tilde g}_{ij}\D z^i\D z^j,\label{Gaussian}
\end{equation}
in which the inverse metric $g^{tt}$ blows up on the horizon defined by $N=0$.

In the coordinate system (\ref{metric-Buchdahl}) or (\ref{metric-Buchdahl-v}), it is easy to identify null infinities.
\begin{lm}
\label{lm:infinity}
In the coordinate system (\ref{metric-Buchdahl}) or (\ref{metric-Buchdahl-v}), $x\to \pm\infty$ are null infinities and a finite $x$ is extendible if it is regular.
\end{lm}
{\it Proof}. 
Consider in the quasi-global coordinates (\ref{metric-Buchdahl}) an affinely parametrized radial null geodesic $\gamma$, of which tangent vector is given by $k^\mu=(\D t/\D\lambda,\D x/\D\lambda,0,\cdots,0)$, where $\lambda$ is an affine parameter along $\gamma$.
Using a conserved quantity $E:=-\xi_\mu k^\mu=Hk^{t}$ along $\gamma$, where $\xi^\mu=(\partial/\partial t)^\mu$ is a Killing vector generating staticity, we write the null condition $k_\mu k^\mu=0$ as $\D x/\D\lambda=\pm E$, which is integrated to give
\begin{align}
x(\lambda)=x_0\pm E\lambda, \label{null-infinity}
\end{align}
where $x_0$ is a constant.
Since $\lambda$ blows up as $x\to \pm \infty$, they are null infinities.
In contrast, a finite $x$ corresponds to a finite $\lambda$, so that it is extendible if it is regular.
\qed

Regularity of the spacetime is required in the definition of the Killing horizon and Lemma~\ref{lm:infinity}.
In fact, how to define regularity depends on the theory under consideration.
In the present paper, we will consider general relativity and Lovelock gravity, of which field equations include up to the second derivatives of the metric.
In such a case, as shown in the following subsections, a $C^{1,1}$ metric is sufficient to avoid curvature singularities as well as the divergence of matter fields.
For this reason, we define the regularity of spacetime by a $C^{1,1}$ metric in the present paper.

\subsection{Curvature singularities}

Curvature singularities are classified into two~\cite{Hawking:1973uf}. 
Although a scalar polynomial (s.p.) curvature singularity is usually examined, it may miss a parallelly propagated (p.p.) curvature singularity.
\begin{dn}[Scalar polynomial curvature singularity]
\label{dn:scalar-sing}
A {\it scalar polynomial (s.p.) curvature singularity} is defined by blowing up of a scalar, formed as a polynomial in the curvature tensor.
\end{dn}
The Ricci scalar $R$ and the Kretschmann scalar $R_{\mu\nu\rho\sigma}R^{\mu\nu\rho\sigma}$ are examples of polynomials in the curvature tensors.
A s.p. curvature singularity is a p.p. curvature singularity but the latter is not always the former. (See Sec.~3 in Ref.~\cite{Ashley:2003tr}.)

\begin{dn}[Parallelly propagated curvature singularity]
\label{dn:pp-sing}
A {\it parallelly propagated (p.p.) curvature singularity} is defined by blowing up of a component of the Riemann tensor $R_{(a)(b)(c)(d)}:=R_{\mu\nu\rho\sigma}E^\mu_{(a)}E^\nu_{(b)}E^\rho_{(c)}E^\sigma_{(d)}$ in a parallelly propagated (pseudo-)orthonormal frame with basis vectors $E^\mu_{(a)}$ along a curve.
\end{dn}
An orthonormal frame is defined by a set of $n$ orthonormal basis vectors $\{{E}^\mu_{(a)}\}$ $(a=0,1,\cdots,n-1)$ that satisfy
\begin{equation}
{E}^\mu_{(a)}{E}_{(b)\mu}=\eta_{(a)(b)}=\mbox{diag}(-1,1,\cdots,1),\label{EE}
\end{equation}
which is equivalent to $g_{\mu\nu}=\eta_{(a)(b)}E^{(a)}_{\mu}E^{(b)}_{\nu}$.
In contrast, basis vectors in a pseudo-orthonormal frame satisfy $g_{\mu\nu}E^\mu_{(a)}E^\nu_{(b)}={\bar \eta}_{(a)(b)}$, where ${\bar \eta}_{(a)(b)}$ is the Minkowski metric in the double-null coordinates given by
\begin{equation} 
\label{flat-null}
{\bar \eta}_{(a)(b)}=\left( 
\vphantom{\begin{array}{c}1\\1\\1\\1\\1\\1\end{array}}
\begin{array}{cccccc}
0 &-1&0&0&\cdots &0\\
-1&0&0&0&\cdots &0\\
0&0&1&0&\cdots&0 \\
0&0&0&\ddots&\vdots&\vdots \\
\vdots&\vdots&\vdots&\cdots&\ddots&0\\
0&0&0 &\cdots&0&1
\end{array}
\right).
\end{equation}
One can choose $E_{(0)}^\mu$ as a tangent vector of an affinely parametrized non-spacelike geodesic $\gamma$ and then $E_{(0)}^\nu\nabla_\nu E_{(0)}^\mu=0$ is satisfied.
In this case, parallelly propagated (pseudo-)orthonormal basis vectors along $\gamma$ satisfy $E_{(0)}^\nu\nabla_\nu E_{(i)}^\mu=0$ for $i=1,2,\cdots,n-1$.

The Kretschmann scalar $K:=R^{\mu\nu}_{~~\rho\sigma}R^{\rho\sigma}_{~~\mu\nu}$ in the quasi-global coordinates (\ref{metric-Buchdahl}) is given by
\begin{align}
K=&4(K_1)^2+4(n-2)(K_2)^2+4(n-2)(K_3)^2 \nonumber\\
&+r^{-4}({}^{(n-2)}{C}^{ij}_{~~kl}{}^{(n-2)}{C}^{kl}_{~~ij})+2(n-2)(n-3)(K_4)^2,\label{K}
\end{align}
where $K_i~(i=1,2,3,4)$ are defined by 
\begin{align}
\label{Ks}
\begin{aligned}
&K_1:=\frac12H'',\qquad K_2:=\frac12H'r^{-1}r',\\
&K_3:=\frac12r^{-1}(2Hr''+r'H'),\qquad K_4:=r^{-2}(k-H{r'}^2).
\end{aligned}
\end{align}
In terms of $K_i$, the Riemann tensor is written as
\begin{align}
\begin{aligned}
R^{tx}_{~~tx}=&-K_1,\qquad {R}^{ti}_{~~tj}=-K_2\delta^i_j,\qquad {R}^{xi}_{~~xj}=-K_3\delta^i_j,\\ 
{R}^{ij}_{~~kl}=&r^{-2}({}^{(n-2)}{C}^{ij}_{~~kl})+K_4(\delta^i_k\delta^j_l-\delta^i_l\delta^j_k).
\end{aligned}
\end{align}
Since we have
\begin{align}
{}^{(n-2)}{C}^{ij}_{~~kl}{}^{(n-2)}{C}^{kl}_{~~ij}={}^{(n-2)}{C}^{(i)(j)}_{~~~~~(k)(l)}{}^{(n-2)}{C}^{(k)(l)}_{~~~~~(i)(j)}\ge 0\label{C2}
\end{align}
due to the Euclidean signature of $K^{n-2}$, the coefficients of each term in the right-hand side of Eq.~(\ref{K}) are non-negative, and hence the divergence of one term implies divergence of $K$.
As a consequence, if ${}^{(n-2)}{C}^{ij}_{~~kl}$ is non-vanishing, $r=0$ corresponds to a s.p. curvature singularity.
\begin{lm}
\label{lm:Weyl}
If the Einstein space $K^{n-2}$ is not maximally symmetric for $n\ge 6$, $r=0$ corresponds to a s.p. curvature singularity.
\end{lm}
{\it Proof}. 
By Eqs.~(\ref{K}) and (\ref{C2}). 
\qed

In the quasi-global coordinates (\ref{metric-Buchdahl}), a very simple criterion is available to identify curvature singularities.
\begin{lm}
\label{lm:finiteness-horizon}
In a spacetime described by the metric (\ref{metric-Buchdahl}) or (\ref{metric-Buchdahl-v}), divergence of $H''$ implies a s.p. curvature singularity (and hence a p.p. curvature singularity), while divergence of $r''$ with $r\ne 0$ implies a p.p. curvature singularity.
\end{lm}
{\it Proof}. 
We introduce the following vectors in the coordinate system (\ref{metric-Buchdahl}):
\begin{align}
&k^\mu\frac{\partial}{\partial x^\mu}=\frac{C}{\sqrt{2}}\biggl(H^{-1}\frac{\partial}{\partial t}-\frac{\partial}{\partial x}\biggl),\\
&l^\mu\frac{\partial}{\partial x^\mu}=\frac{1}{\sqrt{2}C}\biggl(\frac{\partial}{\partial t}+H\frac{\partial}{\partial x}\biggl),\\
&E^\mu_{(i)}\frac{\partial}{\partial x^\mu}=\frac{1}{r}e^j_{(i)}\frac{\partial}{\partial z^j},
\end{align}
where $C$ is a non-zero constant and $e^j_{(i)}$ are basis vectors on $K^{n-2}$ satisfying 
\begin{align}
\gamma_{ij}e_{(k)}^{i}e_{(l)}^{j}=\delta_{(k)(l)}~\Leftrightarrow~\gamma^{ij}=\delta^{(k)(l)}e_{(k)}^{i}e_{(l)}^{j}.\label{e-K}
\end{align}
Here $\gamma^{ij}$ is the inverse of $\gamma_{ij}$.
A null vector $k^\mu$ satisfies $k^\nu \nabla_\nu k^\mu=0$, so that it is a tangent vector of an affinely parametrized radial null geodesic $\gamma$.
Another null vector $l^\mu$ and spacelike vectors $E^\mu_{(i)}$ satisfy $k_\mu l^\mu=-1$ and $g_{\mu\nu}E^\mu_{(i)}E^\nu_{(i)}=\delta_{(i)(j)}$, respectively.
Since $k^\nu \nabla_\nu l^\mu=0$ and $k^\nu \nabla_\nu E_{(i)}^\mu=0$ are satisfied, we identify ${E}^\mu_{(0)}\equiv k^\mu$ and ${E}^\mu_{(1)}\equiv l^\mu$ and then $E^\mu_{(a)}=\{k^\mu, l^\mu,E^\mu_{(i)}\}$ are basis vectors in a parallelly propagated pseudo-orthonormal frame along $\gamma$ satisfying $g_{\mu\nu}E^\mu_{(a)}E^\nu_{(b)}={\bar \eta}_{(a)(b)}$, where ${\bar \eta}_{(a)(b)}$ is given by Eq.~(\ref{flat-null}).
Then, using 
\begin{align}
{R}_{txtx}=&\frac12H'',\quad {R}_{titj}=\frac12HH'rr'\gamma_{ij},\\ 
{R}_{xixj}=&-\frac12r\biggl(2r''+r'\frac{H'}{H}\biggl)\gamma_{ij},\\ 
{R}_{ijkl}=&r^2\left[{}^{(n-2)}{C}_{ijkl}+(k-H{r'}^2)(\gamma_{ik}\gamma _{jl}-\gamma_{il}\gamma _{jk})\right],
\end{align}
we compute non-zero components of $R_{(a)(b)(c)(d)}$ as
\begin{align}
R_{(0)(1)(0)(1)}=&\frac12H'',\label{R(0101)}\\
R_{(0)(i)(0)(j)}=&-\frac{1}{2 r}C^2 r''\delta_{(i)(j)}, \label{R0i0j}\\
R_{(0)(i)(1)(j)}=&R_{(1)(i)(0)(j)}=\frac{1}{2 r}(Hr''+H'r')\delta_{(i)(j)}, \\
R_{(1)(i)(1)(j)}=&-\frac{1}{2 r C^2}H^2 r''\delta_{(i)(j)}, \\
R_{(i)(j)(k)(l)}=&r^{-2}\left[{}^{(n-2)}C_{(i)(j)(k)(l)}+(k-H{r'}^2)(\delta_{(i)(k)}\delta_{(j)(l)}-\delta_{(i)(l)}\delta_{(j)(k)})\right].\label{R(ijkl)}
\end{align} 
Since $R_{(a)(b)(c)(d)}$ are scalars, Eqs.~(\ref{R(0101)})--(\ref{R(ijkl)}) are valid also in the single-null coordinates (\ref{metric-Buchdahl-v}), which share the same $x$ and $z^i$ with the quasi-global coordinates (\ref{metric-Buchdahl}).
Then, by the expression of $K_1$ in Eq. \eqref{Ks} and the Kretschmann scalar \eqref{K}, divergence of $H''$ implies a s.p. curvature singularity (and it is a p.p. curvature singularity by Eq.~(\ref{R(0101)})), while divergence of $r''$ with $r\ne 0$ implies a p.p. curvature singularity by Eq.~(\ref{R0i0j}).
\qed

In Refs.~\cite{Zaslavskii:2007pg,Bronnikov:2008by}, the authors defined a geometric quantity by 
\begin{align}
{\tilde Z}:=H^{-1}(K_2-K_3)=-r^{-1}r'',
\end{align}
where $K_2$ and $K_3$ are given by Eq.~(\ref{Ks}), and classified a surface $x=x_{\rm h}$ as {\it usual} if ${\tilde Z}(x_{\rm h})$ is zero, {\it naked} if it is a finite non-zero value, and {\it truly naked} if it diverges.
By Lemma~\ref{lm:finiteness-horizon}, a {\it truly naked horizon} in Refs.~\cite{Zaslavskii:2007pg,Bronnikov:2008by} is {\it not} a Killing horizon but a curvature singularity.
As a consequence, a {\it truly naked black hole} in Ref.~\cite{Zaslavskii:2007pg} is {\it not} a black hole but, at least, a naked p.p. curvature singularity\footnote{In Ref.~\cite{Bronnikov:2008by}, truly naked horizons are identified as singularities without introducing the concept of p.p. curvature singularity. (See Table~I in Ref.~\cite{Bronnikov:2008by}.)}.

If $x=x_{\rm h}$ is regular, it is a null hypersurface and a Killing horizon by Definition~\ref{dn:K-horizon}.
However, $x=x_{\rm h}$ can be non-null if it is singular.
\begin{lm}
\label{lm:signature}
If $H(x)\propto (x-x_{\rm h})^b$ near $x= x_{\rm h}$ in the coordinate system (\ref{metric-Buchdahl}) or (\ref{metric-Buchdahl-v}), $x=x_{\rm h}$ is causally null for $b\ge 1$ and non-null for $0<b<1$.
\end{lm}
{\it Proof}. 
To clarify the causal nature of spacetime boundaries to draw the Penrose diagram for the metric (\ref{metric-Buchdahl}), we write the two-dimensional Lorentzian portion in the conformally flat form $\D s_2^2=H(x)(-\D t^2+\D x_*^2)$, where $x_*$ is defined by 
\begin{align}
&x_*:=\int\frac{\D x}{H(x)}.\label{def-r*}
\end{align}
Because this two-dimensional metric is conformally flat, $x=x_0$ is causally non-null (null) if it corresponds to finite (infinite) $x_*$.
Thus, if $H(x)\propto (x-x_{\rm h})^b$ near $x= x_{\rm h}$, $x=x_{\rm h}$ is null (non-null) for $b\ge 1$ ($0<b<1$).
\qed

\noindent
By Lemmas~\ref{lm:finiteness-horizon} and~\ref{lm:signature}, $x=x_{\rm h}$ is a non-null curvature singularity for $0<b<1$ (which is timelike in static regions) and a null curvature singularity for $1<b<2$.
In the subtle case of $b=1$, the next leading-order terms in the expansion of $H$ must be taken into account to find if $x=x_{\rm h}$ is singular or not.

\subsection{Regularity with a $C^{1,1}$ metric}

By the contraposition of Lemma~\ref{lm:finiteness-horizon}, the $C^{1,1}$ metric in the coordinates (\ref{metric-Buchdahl}) or (\ref{metric-Buchdahl-v}) is sufficient to avoid p.p. curvature singularities (so scalar polynomial curvature singularities as well).
Namely, a non-singular metric should be at least continuously differentiable ($C^1$) and has the first derivative that is locally Lipschitz continuous, which restricts $H''$ and $r''$ to be finite but allows their finite jumps.

If the metric is not $C^{1,1}$ but just $C^1$, $H'$ and $r'$ are continuous but $H''$ or $r''$ diverges, so that there appears a p.p. curvature singularity.
It is consistent with the statement in Sec.~4.2 in Ref.~\cite{Steinbauer:2022hvq} that the curvature with a $C^{1,1}$ metric is only locally bounded, while the curvature with a $C^1$ metric is merely a distribution of order one and then the geodesic equation fails to be uniquely solvable.
Here it is noted that a $C^1$ metric does not correspond to a thin shell\footnote{A thin shell is also referred to as a {\it massive singular hypersurface} or simply {\it singular hypersurface}.}.
A thin shell is described by a $C^{0,1}$ metric, on which $H$ and $r$ are continuous and $H'$ or $r'$ admits a finite jump, so that it is a curvature singularity but the Riemann tensor can still be treated as the Dirac delta function, which is a distribution of order zero~\cite{Steinbauer:2022hvq}.

For the reasons stated above, in any gravitation theory of which field equations include only up to the second derivatives of the metric such as general relativity and Lovelock gravity, a $C^{1,1}$ metric is sufficient for a regular spacetime and then an energy-momentum tensor $T_{\mu\nu}$ is finite through the field equations.
Accordingly, two solutions described by the metric (\ref{metric-Buchdahl-v}) may be attached at a Killing horizon $x=x_{\rm h}$ without a lightlike thin shell if the metric is $C^{1,1}$ there.
In general relativity, such regularity conditions for the attachment at a null hypersurface are referred to as the Barrab\`{e}s-Israel junction conditions~\cite{Barrabes:1991ng,Poisson:2002nv} for the metric\footnote{If one specifies a matter field, its equation of motion may provide independent junction conditions at a Killing horizon.
(See for example Sec.~4 in Ref.~\cite{Aviles:2019xae} in the case with a scalar field.)}.
(See also Sec.~3.11 in the textbook~\cite{Poissonbook}.)
They consist of the first conditions that require continuity of the induced metric and the second conditions that require continuity of the transverse curvature, which are sufficient to remove the Dirac delta functions that could appear in the field equations. 
As shown in the following proposition, the Barrab\`{e}s-Israel junction conditions are fulfilled if the metric and its first derivative are continuous at the Killing horizon in the coordinates (\ref{metric-Buchdahl-v}).
Then, the metric in the resulting spacetime after the attachment remains $C^{1,1}$ at the horizon as in the two spacetimes to be attached.
\begin{Prop}
\label{prop:attachment}
Consider two solutions described by the metric (\ref{metric-Buchdahl-v}) defined in the domains $x>x_{\rm h}$ and $x<x_{\rm h}$, respectively.
In general relativity, they satisfy the Barrab\`{e}s-Israel junction conditions at a Killing horizon $x=x_{\rm h}$ if $\gamma_{ij}$, $r$, $r'$, and $H'$ are continuous at $x=x_{\rm h}$. 
\end{Prop}
{\it Proof}. 
Let $\Sigma$ be a Killing horizon $x=x_{\rm h}$ in a spacetime described by the metric (\ref{metric-Buchdahl-v}).
The parametric equations $x^\mu=x^\mu({\eta},\theta^i)$ describing $\Sigma$ are $v=\eta$, $x=x_{\rm h}$, and $\theta^i=z^i$.
The line element on $\Sigma$ is $(n-2)$-dimensional and given by
\begin{align}
\D s_{\Sigma}^2=h_{ab}\D w^a \D w^b=r(x_{\rm h})^2\gamma_{ij}\D z^i\D z^j(=\sigma_{ij}\D \theta^i \D 
\theta^j),\label{hab}
\end{align}
where $w^a=(\eta,z^i)$ is a set of coordinates on $\Sigma$.
Tangent vectors of $\Sigma$ defined by $e^\mu_a := \partial x^\mu/\partial w^a$ are given by 
\begin{align}
e^\mu_{\eta}\frac{\partial}{\partial x^\mu}=\frac{\partial}{\partial v},\qquad e^\mu_i\frac{\partial}{\partial 
x^\mu}=\frac{\partial}{\partial z^i}.
\end{align}
We introduce an auxiliary null vector $N^\mu$ given by
\begin{align}
N^\mu \frac{\partial}{\partial x^\mu}=-\frac{\partial}{\partial x} \label{N-attachment}
\end{align}
to complete the basis.
The expression $N_\mu \D x^\mu=-\D v$ shows $N_\mu N^\mu =0$, $N_\mu e^\mu_{\eta}=-1$, and $N_\mu e^\mu_i=0$.
The completeness relation of the basis on $\Sigma$ is given as
\begin{eqnarray}
g^{\mu\nu}|_\Sigma=-k^\mu N^\nu-N^\mu k^\nu+\sigma^{ij} e^\mu_{i}e^\nu_{j}, \label{complete}
\end{eqnarray}
where we have identified $k^\mu\equiv e^\mu_{\eta}$ and $\sigma^{ij}$ is the inverse of $\sigma_{ij}$.

Let $[X]$ be the difference of $X$ evaluated on the two sides of $\Sigma$.
Then, the first Barrab\`{e}s-Israel junction conditions are $[e^\mu_a]=[N^\mu]=[\sigma_{ij}]=0$, while the second Barrab\`{e}s-Israel junction conditions are $N^\rho[\partial_\rho g_{\mu\nu}]=0$, which is equivalent to $[C_{ab}]=0$, where $C_{ab}:=(\nabla_\nu N_{\mu}) e^\mu_{a} e^\nu_b$ is the transverse curvature at $\Sigma$.

In the present case, nonvanishing component of $C_{ab}$ of $\Sigma$ are given by 
\begin{align}
\label{trans-C0}
\begin{aligned}
&C_{\eta\eta}=\Gamma^v_{vv}|_{x=x_{\rm h}}=\frac12H'|_{x=x_{\rm h}},\\
&C_{ij}=\Gamma^v_{ij}|_{x=x_{\rm h}}=-rr'|_{x=x_{\rm h}}\gamma _{ij}.
\end{aligned}
\end{align}
Hence, if $\gamma_{ij}$, $r$, $r'$, and $H'$ are continuous at $x=x_{\rm h}$, the Barrab\`{e}s-Israel junction conditions are satisfied.
\qed

\subsection{Energy-momentum tensors}

In the next section, we will study solutions in general relativity and in Lovelock gravity described by the metric (\ref{metric-Buchdahl}) or (\ref{metric-Buchdahl-v}).
There, orthonormal components of the energy-momentum tensor $T_{(a)(b)}=T_{\mu\nu}E_{(a)}^{\mu}E_{(b)}^{\nu}$ with orthonormal basis vectors $\{{E}^\mu_{(a)}\}$ satisfying Eq.~(\ref{EE}) are generally given by 
\begin{equation} 
\label{T-canonical}
T^{(a)(b)}=\left( 
\vphantom{\begin{array}{c}1\\1\\1\\1\\1\\1\end{array}}
\begin{array}{cccccc}
T^{(0)(0)} &T^{(0)(1)}&0&0&\cdots &0\\
T^{(0)(1)}&T^{(1)(1)}&0&0&\cdots &0\\
0&0&p_2&0&\cdots&0 \\
0&0&0&\ddots&\vdots&\vdots \\
\vdots&\vdots&\vdots&\cdots&\ddots&0\\
0&0&0 &\cdots&0&p_{2}
\end{array}
\right),
\end{equation}
of which components satisfy
\begin{align}
{\cal D}:=(T^{(0)(0)}+T^{(1)(1)})^2-4(T^{(0)(1)})^2\ge 0.\label{det}
\end{align}
By Lemma~1 in Ref.~\cite{Maeda:2022vld}, the Hawking-Ellis type of $T_{\mu\nu}$ is determined as
\begin{align}
\label{D-criterion}
\begin{aligned}
&T^{(0)(1)}=0~~\mbox{or}~~{\cal D}>0~~\Rightarrow~~\mbox{\rm Type~I},\\
&T^{(0)(1)}\ne 0~~\mbox{and}~~{\cal D}=0~~\Rightarrow~~\mbox{\rm Type~II}.
\end{aligned}
\end{align}
We note that, if the inequality~(\ref{det}) is not satisfied, $T^{(a)(b)}$ is of the Hawking-Ellis type IV and violates all the standard energy conditions~\cite{Maeda:2022vld}.

Canonical forms of $T^{(a)(b)}$ are obtained by using a degree of freedom of ${E}^\mu_{(a)}$ provided by the local Lorentz transformation.
(See Sec.~3 in Ref.~\cite{Maeda:2018hqu}.)
The canonical form of type I is given by 
\begin{equation}
\label{T-typeI-0}
T^{(0)(0)}=\mu,\qquad T^{(0)(1)}=0,\qquad T^{(1)(1)}=p_1.
\end{equation}
The standard energy conditions are equivalent to the following inequalities:
\begin{align}
\mbox{NEC}:&~~\mu+p_i\ge 0~~\mbox{for}~~i=1,2,\label{NEC-I}\\
\mbox{WEC}:&~~\mu\ge 0\mbox{~in addition to NEC},\label{WEC-I}\\
\mbox{DEC}:&~~\mu-p_i\ge 0~~\mbox{for}~~i=1,2\mbox{~in addition to WEC},\label{DEC-I}\\
\mbox{SEC}:&~~(n-3)\mu+p_1+(n-2)p_2\ge 0~~\mbox{~in addition to NEC.}\label{SEC-I}
\end{align}
The canonical form of type II is
\begin{equation} 
\label{T-typeII-0}
T^{(0)(0)}=\mu+\nu,\qquad T^{(0)(1)}=\nu,\qquad T^{(1)(1)}=-\mu+\nu
\end{equation}
with $\nu\ne 0$ and the standard energy conditions are equivalent to the following inequalities:
\begin{align}
\mbox{NEC}:&~~\nu\ge 0\mbox{~and~}\mu+p_2\ge 0,\label{NEC-II}\\
\mbox{WEC}:&~~\mu\ge 0\mbox{~in addition to NEC},\label{WEC-II}\\
\mbox{DEC}:&~~\mu-p_2\ge 0\mbox{~in addition to WEC},\label{DEC-II}\\
\mbox{SEC}:&~~(n-4)\mu+(n-2)p_2\ge 0\mbox{~in addition to NEC.}\label{SEC-II}
\end{align}

\section{Killing horizons satisfying $p_{\rm r}\simeq \chi_{\rm r}\rho$ and $p_2\simeq \chi_{\rm t}\rho$}
\label{sec:GR}
In this article, we study Killing horizons in spacetimes described by the metric (\ref{metric-Buchdahl}) or (\ref{metric-Buchdahl-v}) as solutions in general relativity with an Einstein base manifold.
In addition to the gravitational field equations, we will also use the energy-momentum conservation equations $\nabla_\nu T^{\mu\nu}=0$.

\subsection{Non-existence for $\chi_{\rm r}\notin \{-1,[-1/3,0]\}$}

We write the Einstein equations as
\begin{align}
&{\cal G}_{\mu\nu}=\kappa_nT_{\mu\nu},\label{EFE-0}
\end{align}
where ${\cal G}_{\mu\nu}:=G_{\mu\nu}+\Lambda g_{\mu\nu}$.
Non-zero components of ${\cal G}^\mu_{~\nu}$ for the metric (\ref{metric-Buchdahl}) are given by 
\begin{align}
{\cal G}^t_{~t}=&-\frac{n-2}{2}\left[-r^{-1}(H'r' +2Hr'')+(n-3)r^{-2}(k-H{r'}^2)\right]+\Lambda,\\
{\cal G}^x_{~x}=&\frac{n-2}{2}\left[r^{-1}H'r'-(n-3)r^{-2}(k-H{r'}^2)\right]+\Lambda,\\
{\cal G}^i_{~j}=&\biggl\{\frac{1}{2}\left[H''+2(n-3)r^{-1}(H'r'+Hr'')- (n-3)(n-4)r^{-2}(k-H{r'}^2)\right]+\Lambda\biggl\}\delta^i_{~j}.
\end{align}
Hence, the energy-momentum tensor $T_{\mu\nu}$ compatible with the metric (\ref{metric-Buchdahl}) has the following diagonal form
\begin{equation} 
T^{\mu}_{~\nu}=\mbox{diag}(-\rho,p_{\rm r},p_2,\cdots,p_2). \label{T-diagonal}
\end{equation}

In static regions ($H(x)> 0$), we introduce orthonormal basis one-forms as
\begin{align}
\label{basis-diag}
&E^{(0)}_\mu\D x^\mu =\sqrt{H}\D t,\qquad E^{(1)}_\mu\D x^\mu = \sqrt{H^{-1}}\D x,\qquad {E}_\mu^{(k)}\D x^\mu=r{e}_i^{(k)}\D z^i, 
\end{align}
where ${e}_i^{(k)}~(k=2,3,\cdots,n-1)$ are basis one-forms on $K^{n-2}$ satisfying Eq.~(\ref{e-K}).
Then, orthonormal components $T^{(a)(b)}={\cal G}^{\mu\nu}E^{(a)}_{\mu}E^{(b)}_{\nu}/\kappa_n$ are obtained in the type-I form~(\ref{T-typeI-0}) with $\mu=\rho$ and $p_1=p_{\rm r}$, namely,
\begin{equation} 
\label{T-typeI}
T^{(a)(b)}=\mbox{diag}(\rho,p_{\rm r},p_2,\cdots,p_2),
\end{equation}
where $\rho$, $p_{\rm r}$ and $p_2$ are given by 
\begin{align}
\kappa_n\rho=&\frac{n-2}{2}\left[-r^{-1}(H'r' +2Hr'')+(n-3)r^{-2}(k-H{r'}^2)\right]-\Lambda,\label{EFE00}\\
\kappa_np_{\rm r}=&\frac{n-2}{2}\left[r^{-1}H'r'-(n-3)r^{-2}(k-H{r'}^2)\right]+\Lambda,\label{G-xx}\\
\kappa_np_2=&\frac{1}{2}\left[H''+2(n-3)r^{-1}(H'r'+Hr'')- (n-3)(n-4)r^{-2}(k-H{r'}^2)\right]+\Lambda.\label{def-pt}
\end{align}
The quantities $\rho$, $p_{\rm r}$, and $p_2$ are interpreted as the energy density, radial pressure, and tangential pressure of the type-I matter field, respectively.
Equation~(\ref{G-xx}) shows that $p_{\rm r}$ is continuous if the metric is $C^1$.
The dominant energy condition is violated if $|p_{\rm r}/\rho|>1$ or $|p_2/\rho|>1$ is satisfied.

Now Eqs.~(\ref{EFE00}), (\ref{G-xx}), and (\ref{def-pt}) are the Einstein equations with a type-I matter field (\ref{T-diagonal}).
Adding Eqs.~(\ref{EFE00}) and (\ref{G-xx}), we obtain the following simple equation:
\begin{align}
(n-2)r^{-1}Hr''=-\kappa_n(\rho+p_{\rm r}).\label{key}
\end{align}
The energy-momentum conservation equations $\nabla_\mu {T^\mu}_{\nu}=0$ give
\begin{align}
p_{\rm r}'+\frac{(\rho+p_{\rm r})H'}{2H}+\frac{(n-2)r'}{r}(p_{\rm r}-p_2)=0.\label{conserve}
\end{align}
From Eqs.~(\ref{G-xx}), (\ref{key}), and (\ref{conserve}), we obtain the following non-existence result for a Killing horizon.
\begin{Prop}
\label{Prop:non-existence}
Suppose that the components of the energy-momentum tensor (\ref{T-diagonal}) obey $p_{\rm r}\simeq \chi_{\rm r}\rho$ and $p_2\simeq \chi_{\rm t}\rho$ as $x\to x_{\rm h}$, where $\chi_{\rm r}(\ne 0)$ and $\chi_{\rm t}$ are constants.
Then, $\rho$ behaves near $x=x_{\rm h}$ as
\begin{align}
&\rho\simeq \Gamma_0 |H|^{1+\beta}\qquad \biggl(\beta:=-\frac{1+3\chi_{\rm r}}{2\chi_{\rm r}}\biggl), \label{def-beta}
\end{align}
where $\Gamma_0$ is a non-zero constant, and $x=x_{\rm h}$ is at least a p.p. curvature singularity unless $-1/3\le\chi_{\rm r}< 0$ or $\chi_{\rm r}=-1$.
\end{Prop}
{\it Proof}. 
Under the assumptions, Eq.~(\ref{conserve}) gives
\begin{align}
&\chi_{\rm r}\frac{\rho'}{\rho}\simeq -\frac{(1+\chi_{\rm r})H'}{2H}-\frac{(n-2)(\chi_{\rm r}-\chi_{\rm t})r'}{r} \label{conserve2}
\end{align}
near $x=x_{\rm h}$, which is integrated for $\chi_{\rm r}\ne 0$ to give Eq.~(\ref{def-beta}).
Substituting Eq.~(\ref{def-beta}) into Eq.~(\ref{key}), we obtain
\begin{align}
r''\simeq-\frac{\epsilon\kappa_n(1+\chi_{\rm r})r_0\Gamma_0}{n-2} |H|^{\beta} \label{ddr-proof-GR}
\end{align}
near $x=x_{\rm h}$, where $\epsilon:=|H|/H$ and $r_0:=r(x_{\rm h})$.
For $\beta<0$ with $\chi_{\rm r}\ne -1$, Eq.~(\ref{ddr-proof-GR}) diverges as $x\to x_{\rm h}$, so that $x=x_{\rm h}$ is at least a p.p. curvature singularity by Lemma~\ref{lm:finiteness-horizon}.
\qed

\noindent
Here are several remarks on Proposition~\ref{Prop:non-existence}.
\begin{enumerate}

\item The value $\chi_{\rm r}=0$ is excluded in Proposition~\ref{Prop:non-existence}.
In fact, under a stronger condition $p_{\rm r}\equiv 0$ instead of $\lim_{x\to x_{\rm h}}(p_{\rm r}/\rho)\to 0$, static solutions do not admit a Killing horizon because Eq.~(\ref{conserve}) is integrated near $x=x_{\rm h}$ to give
$H\simeq H_0r^{2(n-2)\chi_{\rm t}}$, where $H_0$ is an integration constant, which does not allow $H=0$ with $r> 0$.

\item Proposition~\ref{Prop:non-existence} can be applied to static solutions obeying linear equations of state $p_{\rm r}=\chi_{\rm r}\rho$ and $p_2=\chi_{\rm t}\rho$ and such solutions are studied in Appendix~\ref{app:exact}.
Although it has been claimed in Ref.~\cite{Bronnikov:2008ia} that there is no static solution under a single assumption $p_{\rm r}\equiv 0$ for $k=1$ and $\Lambda=0$, it is true only with $p_2\equiv 0$, namely for a dust fluid, as we give a counterexample in the appendix.

\item Proposition~\ref{Prop:non-existence} explains the transition of a Killing horizon for $n=4$ and $5$ to a p.p. curvature singularity for $n\ge 6$ in the Semiz class-I perfect-fluid solution (\ref{Semiz-I-twopara2}) obeying $p=-(n-3)\rho/(n+1)$~\cite{Semiz:2020lxj,Maeda:2022lsm}.

\end{enumerate}

Proposition~\ref{Prop:non-existence} is a generalization of the result in Ref.~\cite{Bronnikov:2008ia} for $n=4$ and $k=1$ under assumptions $p_{\rm r}=\chi_{\rm r}\rho$, $\rho\ge 0$, and $|p_2|/\rho<\infty$.
Actually, with an additional assumption $\lim_{x\to x_{\rm h}}|H'/H|\to \infty$, one can derive Eq.~(\ref{def-beta}) near a Killing horizon under a weaker condition $\lim_{x\to x_{\rm h}}|p_2/\rho|<\infty$ instead of $p_2\simeq\chi_{\rm t}\rho$ that allows oscillation without convergence.
Under those assumptions, we write Eq.~(\ref{conserve2}) as
\begin{align}
&\chi_{\rm r}\frac{\rho'}{\rho}\simeq -\frac{(1+\chi_{\rm r})H'}{2H}-\frac{(n-2)(\chi_{\rm r}-p_2/\rho)r'}{r} \label{conserve-app}
\end{align}
near $x=x_{\rm h}$.
Since $r$, $r'$, and $H'$ are finite around a Killing horizon, the first term is dominant in the right-hand side if $\chi_{\rm r}\ne -1$ and $\lim_{x\to x_{\rm h}}|H'/H|\to \infty$ are satisfied and then Eq.~(\ref{def-beta}) is obtained for $\chi_{\rm r}\ne 0$.
In this argument, however, it seems to be not so obvious that the additional assumption $\lim_{x\to x_{\rm h}}|H'/H|\to \infty$ always holds for a Killing horizon.

\subsection{Killing horizons for $\chi_{\rm r}=-1$ and $\chi_{\rm r}\in[-1/3,0)$}

Now let us study solutions with a Killing horizon in more detail for $\chi_{\rm r}=-1$ and $\chi_{\rm r}\in[-1/3,0)$.
We first show the following proposition without assuming the linear relations in Proposition~\ref{Prop:non-existence}.
\begin{Prop}
\label{Pro:limit-horizon}
Suppose that $x=x_{\rm h}$ is a Killing horizon.
If $p_{\rm r}\simeq -\rho$ is not satisfied as $x\to x_{\rm h}$, then $\lim_{x\to x_{\rm h}}p_{\rm r}=\lim_{x\to x_{\rm h}}\rho=0$ holds.
If the DEC is assumed in addition, then $\lim_{x\to x_{\rm h}}T_{\mu\nu}=0$ holds.
\end{Prop}
{\it Proof}. 
Because the metric functions $r$ and $H$ are finite up to the second derivatives at a Killing horizon, Eq.~(\ref{key}) shows $\lim_{x\to x_{\rm h}}(p_{\rm r}+\rho)=0$.
Hence, if $p_{\rm r}\simeq -\rho$ is not satisfied as $x\to x_{\rm h}$, then we have $\lim_{x\to x_{\rm h}}p_{\rm r}=\lim_{x\to x_{\rm h}}\rho=0$.
Since the DEC is violated in this case unless $\lim_{x\to x_{\rm h}}p_2=0$, $\lim_{x\to x_{\rm h}}T_{\mu\nu}=0$ holds under the DEC.
\qed

By Proposition~\ref{Pro:limit-horizon}, if a solution admits a Killing horizon under the assumptions in Proposition~\ref{Prop:non-existence} with $\chi_{\rm r}\in[-1/3,0)$, $T_{\mu\nu}$ converges to zero as $x\to x_{\rm h}$. 
However, it does {\it not} necessarily mean that there is no matter field on the Killing horizon because $x=x_{\rm h}$ is a coordinate singularity in the coordinate system (\ref{metric-Buchdahl}).
In fact, although a matter field is of the Hawking-Ellis type I (\ref{T-typeI}) in static regions (and dynamical regions with $H<0$ as well), it is of the Hawking-Ellis type II on Killing horizons if $r''|_{x=x_{\rm h}}\ne 0$ is satisfied.
The following is a trivial generalization of Proposition~2 in Ref.~\cite{Maeda:2021ukk} by adding $\Lambda$.
\begin{Prop}
\label{prop:matter-horizon}
Consider a solution in general relativity described by the metric (\ref{metric-Buchdahl-v}) with a Killing horizon $x=x_{\rm h}$.
If $r''|_{x=x_{\rm h}}=0$ is satisfied, a matter field at $x=x_{\rm h}$ is in the Hawking-Ellis type-I form (\ref{T-typeI}) with $p_{\rm r}=-\rho$, which may be vanishing.
If $r''|_{x=x_{\rm h}}\ne 0$ is satisfied, there exists a non-vanishing matter field of the Hawking-Ellis type II at $x=x_{\rm h}$.
\end{Prop}
{\it Proof}. 
Non-zero components of ${\cal G}^{\mu\nu}$ for the metric (\ref{metric-Buchdahl-v}) are given by 
\begin{align}
{\cal G}^{vv}=&-(n-2)r^{-1}r'',\label{Gvv}\\
{\cal G}^{vx}=&{\cal G}^{xv}=\frac{n-2}{2r^2}\left[rr'H'-(n-3)(k-H{r'}^2)\right]+\Lambda,\\
{\cal G}^{xx}=&H{\cal G}^{vx},\qquad {\cal G}^{ij}=\kappa_np_2r^{-2}\gamma^{ij},\label{Gij-v}
\end{align}
where $p_2$ is given by Eq.~(\ref{def-pt}).
Introducing the following orthonormal basis one-forms
\begin{align}
{E}_\mu^{(0)}\D x^\mu=&-\frac{1}{\sqrt{2}}\left(1+\frac{H}{2}\right)\D v+\frac{1}{\sqrt{2}}\D x,\label{basis0-H}\\
{E}_\mu^{(1)}\D x^\mu=&-\frac{1}{\sqrt{2}}\left(1-\frac{H}{2}\right)\D v-\frac{1}{\sqrt{2}}\D x,\\
{E}_\mu^{(k)}\D x^\mu=&r{e}_i^{(k)}\D z^i,\label{basis2-H}
\end{align}
where basis vectors ${e}_i^{(k)}$ on $K^{n-2}$ satisfy Eq.~(\ref{e-K}), we compute orthonormal components $T^{(a)(b)}={\cal G}^{\mu\nu}E^{(a)}_{\mu}E^{(b)}_{\nu}/\kappa_n$ to give
\begin{align}
\kappa_nT^{(0)(0)}=&\frac{n-2}{8r^2}\left\{-4rr'H'-rr''(H+2)^2+4(n-3)(k-H{r'}^2)\right\}-\Lambda,\label{T00-v}\\
\kappa_nT^{(0)(1)}=&\kappa_nT^{(1)(0)}=\frac{n-2}{8r}r''(H^2-4),\\
\kappa_nT^{(1)(1)}=&\frac{n-2}{8r^2}\left\{4rr'H'-rr''(H-2)^2-4(n-3)(k-H{r'}^2)\right\}+\Lambda,\label{T11-v}\\
\kappa_nT^{(i)(j)}=& \kappa_n\delta^{(i)(j)}p_2,\label{Tij-v}
\end{align}
with which the quantity (\ref{det}) is computed to give
\begin{align}
{\cal D}:=(T^{(0)(0)}+T^{(1)(1)})^2-4(T^{(0)(1)})^2=\frac{(n-2)^2H^2{r''}^2}{\kappa_n^2r^2}.
\end{align}
On the Killing horizon $x=x_{\rm h}$, $T^{(a)(b)}$ is of the Hawking-Ellis type I if $r''|_{x=x_{\rm h}}=0$ and of type II if $r''|_{x=x_{\rm h}}\ne 0$ by the criterion (\ref{D-criterion}).

Since $r''$ is finite at a Killing horizon, $T^{(a)(b)}|_{x=x_{\rm h}}$ is given by 
\begin{equation} 
\label{T-typeII}
T^{(a)(b)}|_{x=x_{\rm h}}=\left( 
\vphantom{\begin{array}{c}1\\1\\1\\1\\1\\1\end{array}}
\begin{array}{cccccc}
\rho_{\rm h}+\nu &\nu&0&0&\cdots &0\\
\nu&-\rho_{\rm h}+\nu&0&0&\cdots &0\\
0&0&p_2|_{x=x_{\rm h}}&0&\cdots&0 \\
0&0&0&\ddots&\vdots&\vdots \\
\vdots&\vdots&\vdots&\cdots&\ddots&0\\
0&0&0 &\cdots&0&p_2|_{x=x_{\rm h}}
\end{array}
\right)
\end{equation}
with
\begin{align}
&\rho_{\rm h}:=\rho|_{x=x_{\rm h}}=\frac{n-2}{2\kappa_n}\left[-r^{-1}r'H'+(n-3)kr^{-2}\right]|_{x=x_{\rm h}} -\frac{\Lambda}{\kappa_n},\label{T00-h}\\
&\nu:=-\frac{n-2}{2\kappa_n}r^{-1}r''|_{x=x_{\rm h}},\label{nu-horizon}\\
&p_2|_{x=x_{\rm h}}=\frac{1}{2\kappa_n}\left[H''+2(n-3)r^{-1}r'H' - (n-3)(n-4)kr^{-2}\right]|_{x=x_{\rm h}}+\frac{\Lambda}{\kappa_n}.\label{Tij-h}
\end{align}
Equation~(\ref{T-typeII}) is a canonical form of the Hawking-Ellis type-II energy-momentum tensor (\ref{T-typeII-0}), which reduces to type I (\ref{T-typeI}) with $p_{\rm r}=-\rho=-\rho_{\rm h}$ if $r''|_{x=x_{\rm h}}=0$ holds.
\qed

It is emphasized that neither Proposition~\ref{Prop:non-existence} nor~\ref{prop:matter-horizon} ensures the existence of static solutions with Killing horizons for $\chi_{\rm r}\in[-1/3,0)$ or $\chi_{\rm r}=-1$.
Nevertheless, for $\chi_{\rm r}=-1$, the electrically charged Reissner-Nordstr\"om solution is a typical example of static solutions that admits Killing horizons, for which $p_{\rm r}/\rho=-1$ and $p_2/\rho=1$ are satisfied everywhere.
The following proposition explains properties of more general solutions for $\chi_{\rm r}=-1$.
\begin{Prop}
\label{Prop:chi=-1}
Suppose that the Einstein equations admit non-vacuum solutions with a Killing horizon described by the metric (\ref{metric-Buchdahl-v}) with non-constant $r$ under the assumptions in Proposition~\ref{Prop:non-existence} with $\chi_{\rm r}=-1$.
Then, the solution near the horizon is given by Eq.~(\ref{chi=-1-sol1}) for $\chi_{\rm t}\ne 1/(n-2)$ and Eq.~(\ref{chi=-1-sol1-special}) for $\chi_{\rm t}= 1/(n-2)$ with the same values of the parameters $M$ and $\eta$ on both sides of the horizon.
On the horizon, there exists a matter field of the Hawking-Ellis type~I.
\end{Prop}
{\it Proof}.
Under exact relations $p_{\rm r}=-\rho$ and $p_2=\chi_{\rm t}\rho$, the general static solution with non-constant $r$ is given by Eq.~(\ref{chi=-1-sol1}) for $\chi_{\rm t}\ne 1/(n-2)$ and Eq.~(\ref{chi=-1-sol1-special}) for $\chi_{\rm t}= 1/(n-2)$, which approximate more general static solutions when $p_{\rm r}\simeq -\rho$ and $p_2\simeq \chi_{\rm t}\rho$ are satisfied.
The solution is parametrized by $M$ and $\eta$ and admits a Killing horizon in certain parameter regions.
The equation $H(x_{\rm h})=0$ gives
\begin{align}
\label{Mh}
2M= 
\begin{dcases}
kx_{\rm h}^{n-3}+\eta x_{\rm h}^{1-(n-2)\chi_{\rm t}}-\frac{2\Lambda x_{\rm h}^{n-1}}{(n-1)(n-2)} & [\chi_{\rm t}\ne 1/(n-2)]\\
kx_{\rm h}^{n-3}+\eta \ln|x_{\rm h}|-\frac{2\Lambda x_{\rm h}^{n-1}}{(n-1)(n-2)} & [\chi_{\rm t}= 1/(n-2)]
\end{dcases}
,
\end{align}
with which we obtain
\begin{equation}
\label{chi=-1-sol1-H'-horizon}
H'(x_{\rm h})= 
\begin{dcases}
\frac{(n-3)k}{x_{\rm h}}-\frac{[(n-2)\chi_{\rm t}-1]\eta}{x_{\rm h}^{(n-2)\chi_{\rm t}+(n-3)}}-\frac{2\Lambda x_{\rm h}}{n-2} & [\chi_{\rm t}\ne 1/(n-2)]\\
\frac{(n-3)k}{x_{\rm h}}+\frac{\eta}{x^{n-2}}-\frac{2\Lambda x_{\rm h}}{n-2} & [\chi_{\rm t}= 1/(n-2)]
\end{dcases}
.
\end{equation}
Let $(M,\eta)=(M_+,\eta_+)$ and $(M,\eta)=(M_-,\eta_-)$ be the values of the parameters in the regions $x>x_{\rm h}$ and $x<x_{\rm h}$, respectively.
Since $k$, $\Lambda$, and $\chi_{\rm t}$ are the same on both sides of the horizon, the continuity of $H'(x)$ at $x=x_{\rm h}$ requires $\eta_+=\eta_-$ and then $M_+=M_-$ holds by Eq.~(\ref{Mh}).
Since $r''(x_{\rm h})=0$ and $\rho(x_{\rm h})\ne 0$ are satisfied, there exists a matter field on the horizon which is of the Hawking-Ellis type~I by Proposition~\ref{prop:matter-horizon}.
\qed

Now we show that such static solutions with a Killing horizon certainly exist also for $\chi_{\rm r}\in[-1/3,0)$.

\begin{Prop}
\label{Prop:asymp}
Under the assumptions in Proposition~\ref{Prop:non-existence} with $\chi_{\rm r}\in[-1/3,0)$, the Einstein equations admit non-vacuum solutions with a non-degenerate Killing horizon.
For $\chi_{\rm r}\ne -1/(1+2N)$ with $N\in\mathbb{N}$, the metric (\ref{metric-Buchdahl-v}) is not $C^{[2+\beta],1}$ but $C^{[2+\beta]}$ on the horizon, where $\beta$ is defined by Eq.~(\ref{def-beta}). 
For $\chi_{\rm r}= -1/(1+2N)$, the metric is at least $C^{2+\beta}$ on the horizon if the parameter $r_{2+\beta}$ in the asymptotic expansion of $r(x)$, given by Eq.~(\ref{r-asymp2-special-out1}) for $(n-3)k=\Lambda=0$ and by Eq.~(\ref{r-asymp2}) otherwise, takes the same value on both sides of the horizon.
If not, the metric is not $C^{2+\beta}$ but $C^{1+\beta,1}$ on the horizon.
\end{Prop}
{\it Proof}.
Basic equations are given by 
\begin{align}
r''=&-\frac{\kappa_n(p_{\rm r}+\rho)r}{(n-2)H},\label{beq-asymp1}\\
H'= &\frac{(n-2)(n-3)(k-H{r'}^2)+2\kappa_np_{\rm r} r^2-2\Lambda r^2}{(n-2)rr'},\label{beq-asymp2}\\
0=&p_{\rm r}'+\frac{(\rho+p_{\rm r})H'}{2H}+\frac{(n-2)r'}{r}(p_{\rm r}-p_2),\label{conserve-asymp}\\
\kappa_np_2=&\frac{1}{2r^2}\left\{H''r^2+2(n-3)r(H' r'+Hr'')- (n-3)(n-4)(k-H{r'}^2)\right\}+\Lambda,\label{beq-asymp3}
\end{align}
which are Eqs.~(\ref{key}), (\ref{G-xx}), (\ref{conserve}), and (\ref{def-pt}), respectively.
We perform an asymptotic analysis near a non-degenerate Killing horizon $x=x_{\rm h}$ with $r(x_{\rm h})=r_0(>0)$, where $H(x_{\rm h})=0$ and $H'(x_{\rm h})\ne 0$ hold.
The details are shown in Appendix~\ref{app:coefficients}.

Under the assumptions in Proposition~\ref{Prop:non-existence} with $\chi_{\rm r}\in[-1/3,0)$, Eq.~(\ref{conserve}) gives Eq.~(\ref{def-beta}) with $\beta\in[0,\infty)$, where $\beta=0$ corresponds to $\chi_{\rm r}=-1/3$.
To be regular at $x=x_{\rm h}$, the metric functions $r$ and $H$ must be finite up to the second derivative by Lemma~\ref{lm:finiteness-horizon}.
Then, from Eqs.~(\ref{beq-asymp1}), (\ref{beq-asymp2}), and (\ref{beq-asymp3}) with Eq.~(\ref{def-beta}), we respectively obtain
\begin{align}
&r''|_{x=x_{\rm h}}=-\frac{\epsilon\kappa_n(1+\chi_{\rm r})}{n-2}\Gamma_0r_0 |H|^{\beta}|_{x=x_{\rm h}},\label{r''-H}\\
&rr'H'|_{x=x_{\rm h}}=(n-3)k-\frac{2\Lambda r_0^2}{n-2},\label{rr'H'}\\
&r^2H''|_{x=x_{\rm h}}=-(n-2)(n-3)k+\frac{2(n-4)\Lambda}{n-2} r_0^2,\label{const1}
\end{align}
where $\epsilon:=|H|/H$.

First we consider the case $\beta\in\mathbb{N}_0$, or equivalently $\chi_{\rm r}= -1/(1+2N)$ with $N\in \mathbb{N}$.
In the case where at least $(n-3)k$ or $\Lambda$ is non-zero, the asymptotic solution near $x=x_{\rm h}$ is given by 
\begin{align}
&H(x)\simeq \sum_{i=1}^{3+\beta}H_i\Delta^i,\label{H-asymp2}\\
&r(x)\simeq r_0+r_1\Delta+r_{2+\beta}\Delta^{2+\beta},\label{r-asymp2}\\
&\rho\simeq \rho_{1+\beta}\Delta^{1+\beta},\label{rho-asymp2-20}
\end{align}
where $\Delta:=x-x_{\rm h}$ and 
\begin{align}
&\rho_{1+\beta}=-\frac{(n-2)(2+\beta)(1+\beta)H_1r_{2+\beta}}{\kappa_n(1+\chi_{\rm r})r_0}.
\end{align}
The parameters of the solution are $r_0(> 0)$, $r_1$ (or $H_1$), and $r_{2+\beta}$ and the coefficients $H_i$ are given such as
\begin{align}
&H_1=\frac{(n-2)(n-3)k-2\Lambda {r_0}^2}{(n-2)r_0r_1},\label{H1-out}\\
&H_2=\frac{-(n - 2)^2(n - 3)k + 2(n - 4)\Lambda {r_0}^2}{2(n-2){r_0}^2},\label{H2-out}\\
&H_3=\frac{(n-3)r_1\{(n-1)(n - 2)k - 2\Lambda {r_0}^2\}}{6{r_0}^3}\quad (\mbox{for}~\beta\ge 1),\label{H3-out}\\
&H_{i+1}=-\frac{(n-3+i)r_1}{(1+i)r_0}H_{i}\quad (\mbox{with}~3\le i\le [1+\beta]~\mbox{for}~\beta\ge 2),\label{H-recursion}
\end{align}
where $[1+\beta]$ is the integer part of $1+\beta$.
The last coefficient $H_{3+\beta}$ is given by
\begin{align}
H_{3}=&\frac{(n-3)r_1\left\{(n-1)(n-2)k-2\Lambda {r_0}^2\right\}}{6{r_0}^3}-\frac{2H_1r_{2}}{3r_0}\biggl\{\frac{(n-2)\chi_{\rm t}}{1+\chi_{\rm r}} +2(n-3)\biggl\}
\end{align}
for $\beta=0$, 
\begin{align}
H_{4}=&-\frac{n(n-3){r_1}^2\left\{(n-1)(n-2)k-2\Lambda {r_0}^2\right\}}{24{r_0}^4}-\frac{H_1r_{3}}{2r_0}\biggl\{\frac{2(n-2)\chi_{\rm t}}{1+\chi_{\rm r}} +3(n-3)\biggl\}
\end{align}
for $\beta=1$, and 
\begin{align}
H_{3+\beta}=&\frac{{r_1}^2(n-1+\beta)(n-2+\beta)}{{r_0}^2(3+\beta)(2+\beta)}H_{1+\beta} \nonumber\\
&-\frac{2H_1r_{2+\beta}}{(3+\beta)r_0}\biggl\{\frac{(1+\beta)(n-2)\chi_{\rm t}}{1+\chi_{\rm r}} +(2+\beta)(n-3)\biggl\} \label{H3+beta-N}
\end{align}
for $\beta\ge 2$, where $H_{1+\beta}$ on the right-hand side can be written in terms of $r_0$ and $r_1$ by Eqs.~(\ref{H3-out}) and (\ref{H-recursion}).

For $(n-3)k=\Lambda=0$, the asymptotic solution is given by
\begin{align}
&H(x)\simeq H_1\Delta+H_{3+\beta}\Delta^{3+\beta},\label{H-asymp2-special-out1}\\
&r(x)\simeq r_0+r_{2+\beta}\Delta^{2+\beta}\label{r-asymp2-special-out1}
\end{align}
and Eq.~(\ref{rho-asymp2-20}), where $H_{3+\beta}$ is given by 
\begin{align}
&H_{3+\beta}=-\frac{2H_1r_{2+\beta}}{(3+\beta)r_0}\biggl\{\frac{(1+\beta)(n-2)\chi_{\rm t}}{1+\chi_{\rm r}} +(2+\beta)(n-3)\biggl\}.\label{H3+beta-out-main}
\end{align}
In this case, the parameters of the solution are $r_0$, $r_{2+\beta}$, and $H_1(\ne 0)$.

For $\beta\in\mathbb{N}_0$, the continuity of the metric (\ref{metric-Buchdahl-v}) and its first derivative requires that $r_0$ and $r_1$ (or $H_1$) in Eqs.~(\ref{H-asymp2}) and (\ref{r-asymp2}) and $r_0$ and $H_1$ in Eqs.~(\ref{H-asymp2-special-out1}) and (\ref{r-asymp2-special-out1}) are the same on both sides of the horizon $x=x_{\rm h}$. 
The metric is at least $C^{2+\beta}$ on the horizon if $r_{2+\beta}$ takes the same value on both sides of the horizon.
If not, the $(2+\beta)$-th derivative of $r$ is finite but discontinuous on the horizon and the metric is not $C^{2+\beta}$ but $C^{1+\beta,1}$ there.

For a positive non-integer $\beta$, or equivalently $\chi_{\rm r}\ne -1/(1+2N)$ with $N\in \mathbb{N}$, the asymptotic solutions around the horizon $x=x_{\rm h}$ are given in the domains $x\ge x_{\rm h}$ and $x\le x_{\rm h}$ separately.
In the case where at least $(n-3)k$ or $\Lambda$ is non-zero, the asymptotic solution near $x=x_{\rm h}$ in the domain of $x\ge x_{\rm h}$ is given by 
\begin{align}
&H(x)\simeq \sum_{i=1}^{[3+\beta]}H_i\Delta^i+H_{3+\beta}\Delta^{3+\beta},\label{H-asymp2-non}\\
&r(x)\simeq r_0+r_1\Delta+r_{2+\beta}\Delta^{2+\beta},\label{r-asymp2-non}
\end{align}
and $\rho$ given by Eqs.~(\ref{rho-asymp2-20}).
The parameters of the solution are $r_0(> 0)$, $r_1$ (or $H_1$), and $r_{2+\beta}$ and $H_i$ are determined as Eqs.~(\ref{H1-out})--(\ref{H-recursion}) and Eq.~(\ref{H3+beta-out-main}). The last coefficient $H_{[3+\beta]}$ is given by
\begin{align}
&H_3(=H_{[3+\beta]})=\frac{(n-3)r_1\{(n-1)(n - 2)k - 2\Lambda {r_0}^2\}}{6{r_0}^3}\label{H3-prop}
\end{align}
for $0<\beta<1$, 
\begin{align}
H_4(=H_{[3+\beta]})=&-\frac{n(n-3){r_1}^2\{(n-1)(n - 2)k - 2\Lambda {r_0}^2\}}{24{r_0}^4}=-\frac{nr_1}{4r_0}H_3\label{H4-prop}
\end{align}
for $1<\beta<2$, and 
\begin{align}
H_{[3+\beta]}=&-\frac{2r_1[n-2+\beta]}{r_0[3+\beta]}H_{[2+\beta]} -\frac{{r_1}^2\{(n-3)(n-4+2[1+\beta])+[1+\beta][\beta]\}}{{r_0}^2[3+\beta][2+\beta]}H_{[1+\beta]}\label{H[3+beta]-out-main}
\end{align}
for $\beta>2$.
On the other hand, the asymptotic solution in the domain $x\le x_{\rm h}$ is given by 
\begin{align}
&H(x)\simeq \sum_{i=1}^{[3+\beta]}H_i\Delta^i +{\bar H}_{3+\beta}(-\Delta)^{3+\beta},\label{expansion-H1-in}\\
&r(x)\simeq r_0+r_1\Delta+{\bar r}_{2+\beta}(-\Delta)^{2+\beta},\label{expansion-r1-in}\\
&\rho\simeq {\bar \rho}_{1+\beta}(-\Delta)^{1+\beta},\label{expansion-rho1-in}
\end{align}
where 
\begin{align}
&{\bar \rho}_{1+\beta}=\frac{(n-2)(2+\beta)(1+\beta)H_1{\bar r}_{2+\beta}}{\kappa_n(1+\chi_{\rm r})r_0}.
\end{align}
Here $r_0(>0)$, $r_1$ (or $H_1$), and ${\bar r}_{2+\beta}$ are parameters and the coefficients $H_i$ and ${\bar H}_{3+\beta}$ are determined as Eqs.~(\ref{H1-out})--(\ref{H-recursion}), Eqs.~(\ref{H3-prop})--(\ref{H[3+beta]-out-main}), and
\begin{align}
{\bar H}_{3+\beta}=\frac{2H_1{\bar r}_{2+\beta}}{(3+\beta)r_0}\biggl\{\frac{(1+\beta)(n-2)\chi_{\rm t}}{1+\chi_{\rm r}} +(2+\beta)(n-3)\biggl\}.\label{barH3+beta-in-main}
\end{align}

For $(n-3)k=\Lambda=0$, the asymptotic solution near $x=x_{\rm h}$ in the domain of $x\ge x_{\rm h}$ is given by 
\begin{align}
&H(x)\simeq H_1\Delta+H_{3+\beta}\Delta^{3+\beta},\label{H-asymp2-special-out2}\\
&r(x)\simeq r_0+r_{2+\beta}\Delta^{2+\beta}\label{r-asymp2-special-out2}
\end{align}
with Eq.~(\ref{rho-asymp2-20}).
The parameters of the solution are $r_0$, $r_{2+\beta}$, and $H_1(\ne 0)$ and $H_{3+\beta}$ is given by Eq.~(\ref{H3+beta-out-main}).
The asymptotic solution in the domain of $x\le x_{\rm h}$ is given by 
\begin{align}
&H(x)\simeq H_1\Delta+{\bar H}_{3+\beta}(-\Delta)^{3+\beta},\label{expansion-H1-in-special}\\
&r(x)\simeq r_0+{\bar r}_{2+\beta}(-\Delta)^{2+\beta}\label{expansion-r1-in-special}
\end{align}
with Eq.~(\ref{expansion-rho1-in}).
The parameters in this domain are $r_0(>0)$, ${\bar r}_{2+\beta}$, and $H_1(\ne 0)$ and ${\bar H}_{3+\beta}$ is given by Eq.~(\ref{barH3+beta-in-main}).

For a non-integer positive $\beta$, the continuity of the metric (\ref{metric-Buchdahl-v}) and its first derivative requires that $r_0$ and $r_1$ (or $H_1$) in Eqs.~(\ref{H-asymp2-non}), (\ref{r-asymp2-non}), (\ref{expansion-H1-in}), and (\ref{expansion-r1-in}) and $r_0$ and $H_1$ in Eqs.~(\ref{H-asymp2-special-out2})--(\ref{expansion-r1-in-special}) are the same on both sides of the horizon $x=x_{\rm h}$. 
Then, for $r_{2+\beta}\ne 0$ or ${\bar r}_{2+\beta}\ne 0$, the $[3+\beta]$-th derivative of $r$ blows up as $x\to x_{\rm h}$ at least on one side of the horizon, so that the metric is not $C^{[2+\beta],1}$ but $C^{[2+\beta]}$ on the horizon.
\qed

\noindent
Here are several remarks on Proposition~\ref{Prop:asymp}.
\begin{enumerate}

\item Without loss of generality, $x_{\rm h}$ can be set to any finite value by a shift transformation $x\to x+x_0$ with $x_0\in \mathbb{R}$.
Moreover, the metric (\ref{metric-Buchdahl}) is invariant under coordinate transformations $t\to \omega t$ and $x\to x/\omega$ and a redefinition $\omega^2 H\to H$ with a real non-zero constant $\omega$.
Therefore, the asymptotic solutions are two parameter families.

\item Since the metric is at least $C^{1,1}$, a solution in the domain $x>x_{\rm h}$ and a solution in the domain $x<x_{\rm h}$ can be attached at the Killing horizon in a regular manner without a lightlike thin shell.
As a special case, one can attach to the Schwarzschild-Tangherlini-type vacuum solution which is realized for $r_{2+\beta}=0$ or ${\bar r}_{2+\beta}=0$.

\item For $H_1>(<)0$, the region of $x> x_{\rm h}$ is static (dynamical) and the region of $x< x_{\rm h}$ is dynamical (static).

\item In a static region, the NEC is violated if $\rho<0$.
In a dynamical region, where $t$ and $x$ are spacelike and timelike coordinates, respectively, the energy density $\mu$ and the radial pressure $p_1$ are given by $\mu=-p_{\rm r}$ and $p_1=-\rho$, respectively, so that the NEC is violated if $\rho> 0$ is satisfied in a dynamical region.

\end{enumerate}

Proposition~\ref{Prop:asymp} asserts the existence of non-degenerate horizons for $\chi_{\rm r}\in[-1/3,0)$.
The following proposition shows that degenerate horizons are allowed only for $(n-3)k\Lambda>0$.

\begin{Prop}
\label{Prop:degenerate}
Under the assumptions in Proposition~\ref{Prop:non-existence} with $\chi_{\rm r}\in[-1/3,0)$, the Einstein equations admit non-vacuum solutions described by the metric (\ref{metric-Buchdahl}) or (\ref{metric-Buchdahl-v}) with a degenerate Killing horizon only for $(n-3)k\Lambda>0$ and then $r(x_{\rm h})=\sqrt{(n-2)(n-3)k/(2\Lambda)}$ and $H''(x_{\rm h})\ne 0$ is satisfied.
\end{Prop}
{\it Proof}.
By Eqs.~(\ref{r''-H})--(\ref{const1}), degenerate horizons are allowed only for $(n-3)k\Lambda>0$ and the asymptotic form of the solution near the horizon is given by 
\begin{align}
&H(x)\simeq \sum_{i=2}^{[4+2\beta]}H_i\Delta^i+(\beta-[\beta])H_{4+2\beta}\Delta^{4+2\beta},\label{H-asymp2-app}\\
&r(x)\simeq r_0+r_1\Delta+r_{2+2\beta}\Delta^{2+2\beta},\label{r-asymp2-app}\\
&\rho\simeq \rho_{2+2\beta} \Delta^{2+2\beta} \label{rho-asymp2-app}
\end{align}
with 
\begin{align}
\label{Crho-app2}
\begin{aligned}
&H_2=-\frac{2\Lambda}{n-2}(\ne 0),\qquad r_0^2=\frac{(n-2)(n-3)k}{2\Lambda},\\
&\rho_{2+2\beta}=-\frac{(n-2)(2+2\beta)(1+2\beta)H_2r_{2+2\beta}}{\kappa_n(1+\chi_{\rm r})r_0}.
\end{aligned}
\end{align}
\qed

Lastly, we show that there exists a matter field on the Killing horizon only for $\chi_{\rm r}=-1/3$.

\begin{Prop}
\label{Prop:matter-horizon}
Under the assumptions in Proposition~\ref{Prop:non-existence} with $\chi_{\rm r}\in[-1/3,0)$, there exists a matter field on the Killing horizon $x=x_{\rm h}$ only for $\chi_{\rm r}=-1/3$.
Orthonormal components of the energy-momentum tensor on the horizon are given in the Hawking-Ellis type-II form~(\ref{T-typeII}) with $\rho_{\rm h}=p_2|_{x=x_{\rm h}}=0$ and $\nu=-(n-2)r_2/(\kappa_nr_0)\ne 0$.
\end{Prop}
{\it Proof}.
Substituting Eqs.~(\ref{r''-H})--(\ref{const1}) into Eqs.~(\ref{T00-h})--(\ref{Tij-h}), we obtain $T^{(a)(b)}|_{x=x_{\rm h}}$ in the type II form (\ref{T-typeII}) with 
\begin{align}
&\rho_{\rm h}=p_2|_{x=x_{\rm h}}=0,\qquad \nu=\frac{1}{2}\epsilon(1+\chi_{\rm r})\Gamma_0 |H|^{\beta}|_{x=x_{\rm h}}.\label{matter-on-horizon1-GR}
\end{align}
For $\chi_{\rm r}\in[-1/3,0)$, we have $\beta\in[0,\infty)$ and hence $\nu\ne 0$ is satisfied only for $\chi_{\rm r}=-1/3$.
Substituting Eqs.~(\ref{r-asymp2}) and (\ref{r-asymp2-app}) with $\beta=0$ for a non-degenerate horizon and a degenerate horizon, respectively, into Eq.~(\ref{nu-horizon}), we obtain
\begin{align}
&\nu=-\frac{(n-2)r_2}{\kappa_nr_0}
\end{align}
in both cases.
\qed

\noindent
The matter field on the Killing horizon for $\chi_{\rm r}=-1/3$ in Proposition~\ref{Prop:matter-horizon} can be interpreted as a null dust fluid, of which energy-momentum tensor is given by 
\begin{align}
&T^{\mu\nu}|_{x=x_{\rm h}}=\rho k^\mu k^\nu,\qquad k^\mu\frac{\partial}{\partial x^\mu}=k^{v}\frac{\partial}{\partial v},\label{null-dust}
\end{align}
where the energy density $\rho$ and a null vector $k^\mu$ satisfy $\rho(k^{v})^2=2\nu$, where $\nu$ is given by Eq.~(\ref{matter-on-horizon1-GR}).
Since a null dust is equivalent to a massless scalar field with null gradient~\cite{Faraoni:2018fil}, one might naively expect that the matter field on the horizon might also be interpreted as the following massless scalar field:
\begin{align}
\label{phi-H}
\begin{aligned}
&T^{\mu\nu}|_{x=x_{\rm h}}=(\nabla^\mu\phi)(\nabla^\nu\phi)-\frac12g^{\mu\nu}(\nabla\phi)^2,\\
&\phi=\phi_{\rm h}\pm \sqrt{2\nu} \Theta(x-x_{\rm h}),
\end{aligned}
\end{align}
where $\phi_{\rm h}$ is a constant and $\Theta(x)$ is the Heaviside step function.
However, it is not true firstly because $\D\phi/\D x$ contains the Dirac delta function $\delta(x-x_{\rm h})$, so that $T^{\mu\nu}|_{x=x_{\rm h}}$ contains its square and hence it is not well-defined.
The second reason is that the scalar field (\ref{phi-H}) is not continuous at $x=x_{\rm h}$ and does not satisfy the first junction conditions in the system with a minimally coupled scalar field~\cite{Aviles:2019xae}.

In Propositions~\ref{Prop:chi=-1} and \ref{Prop:asymp}, linear relations $p_{\rm r}\simeq \chi_{\rm r}\rho$ and $p_2\simeq \chi_{\rm t}\rho$ are assumed near $x=x_{\rm h}$ with $\chi_{\rm r}=-1$ and $\chi_{\rm r}\in[-1/3,0)$, respectively.
If we assume those relations everywhere, not just near the horizon, the metric is analytic on the horizon for $\chi_{\rm r}=-1$ and can be $C^{\infty}$ there only for $\chi_{\rm r}= -1/(1+2N)$ with $N\in\mathbb{N}$, as shown below.
While non-analytic extensions beyond the horizon are not allowed for $N=0$ ($\chi_{\rm r}=-1$), they are possible for $N\in\mathbb{N}$.
\begin{Prop}
\label{Prop:analytic-metric}
Suppose that a static solution described by the metric (\ref{metric-Buchdahl}) with a non-constant $r$ admits a Killing horizon in general relativity with an energy-momentum tensor (\ref{T-diagonal}) obeying $p_{\rm r}=\chi_{\rm r}\rho$ and $p_2=\chi_{\rm t}\rho$ with $\chi_{\rm r}=-1$ or $\chi_{\rm r}\in[-1/3,0)$.
Then, for $\chi_{\rm r}=-1$, the metric (\ref{metric-Buchdahl-v}) is analytic ($C^\omega$) on the horizon.
For $\chi_{\rm r}\in[-1/3,0)$, the metric can be $C^{\infty}$ on the horizon only for $\chi_{\rm r}= -1/(1+2N)$ with $N\in\mathbb{N}$.
\end{Prop}
{\it Proof}.
For $\chi_{\rm r}=-1$, the solution is given by the metric (\ref{metric-Buchdahl}) with Eq.~(\ref{chi=-1-sol1}) for $\chi_{\rm t}\ne 1/(n-2)$ and Eq.~(\ref{chi=-1-sol1-special}) for $\chi_{\rm t}= 1/(n-2)$.
Since the parameters $M$ and $\eta$ are the same on both sides of the horizon by Proposition~\ref{Prop:chi=-1}, the metric (\ref{metric-Buchdahl-v}) is analytic on the horizon.

For $\chi_{\rm r}\in[-1/3,0)$, the metric cannot be $C^{\infty}$ on the horizon except for $\chi_{\rm r}= -1/(1+2N)$ with $N\in\mathbb{N}$, which is equivalent to $\beta\in\mathbb{N}_0$, by Proposition~\ref{Prop:asymp}.
In that case, if one chooses the same value of the parameter $r_{2+\beta}$ on both sides of the horizon in the asymptotic solution, given by Eqs.~(\ref{H-asymp2-special-out1}) and (\ref{r-asymp2-special-out1}) for $(n-3)k=\Lambda=0$ and Eqs.~(\ref{H-asymp2}) and (\ref{r-asymp2}) otherwise, the metric (\ref{metric-Buchdahl-v}) is at least $C^{2+\beta}$ on the horizon.
However, even when higher-order terms are taken into account in the asymptotic expansion, non-integer powers do not appear and therefore the metric is $C^{\infty}$ on the horizon.
\qed

\subsection{Comparison with the results in Ref.~\cite{Bronnikov:2008ia}}
\label{sec:comparison}

Propositions~\ref{Prop:chi=-1} and \ref{Prop:asymp} are generalizations of the result in Ref.~\cite{Bronnikov:2008ia} for $n=4$ and $k=1$ under assumptions $p_{\rm r}\simeq \chi_{\rm r}\rho$, $\rho\ge 0$, and $|p_2|/\rho<\infty$ near $x=x_{\rm h}$.
The last assumption is weaker than our assumption $p_2\simeq \chi_{\rm t}\rho$ in Proposition~\ref{Prop:non-existence}.
The correspondence between the present paper and Ref.~\cite{Bronnikov:2008ia} is shown in Table~\ref{table:correspondence}.
\begin{table*}[htb]
\begin{center}
\caption{Correspondence between the present paper with $n=4$ and $k=1$ and Ref.~\cite{Bronnikov:2008ia}.}
\label{table:correspondence}
\begin{tabular}{|c|c|c|c|}\hline
& Present paper & Ref.~\cite{Bronnikov:2008ia} \\ \hline
Signature & $(-,+,+,+)$ & $(+,-,-,-)$ \\ \cline{1-3}
Radial coordinate & $x$ & $u$ \\ \cline{1-3}
Metric functions & $H(x)$ & $A(u)$ \\ \cline{2-3}
& $r(x)$ & $r(u)$ \\ \hline
Matter functions & $\rho$ & $\rho$ \\ \cline{2-3}
& $p_{\rm r}$ & $p_{\rm r}$ \\ \cline{2-3}
& $p_2$ & $p_{\perp}$ \\ \cline{2-3}
& $\Lambda$ & $\rho_{\rm (vac)}$ \\ \cline{2-3}
& $\Lambda$ & $\rho_{\rm r(vac)}$ \\ \cline{2-3}
& $-\Lambda$ & $p_{\rm \perp(vac)}$ \\ \hline
Constants & $\kappa_4(=8\pi G_4)$ & $8\pi$ \\ \cline{2-3}
& $\chi_{\rm r}$ & $w$ \\ \cline{2-3}
& $\beta$ & $-(1+3w)/(2w)$ \\ \hline
\end{tabular}
\end{center}
\end{table*}

Theorem~1 in Ref.~\cite{Bronnikov:2008ia} is a result in the case where ${\rho_{\rm (vac)}}$, $\rho_{\rm r(vac)}$, and ${p_{\rm \perp(vac)}}$ are vanishing, which corresponds to $\Lambda=0$ in the present paper.
It asserts that static solutions admit Killing horizons only for $\chi_{\rm r}=-1$ and $\chi_{\rm r}= -1/(1+2{\cal K})$ with ${\cal K}\in\mathbb{N}$, where ${\cal K}$ corresponds to $k$ in Ref.~\cite{Bronnikov:2008ia}.
However, analyticity of the metric is implicitly assumed in that theorem.
Proposition~\ref{Prop:asymp} in the present paper asserts that Killing horizons described by a non-analytic metric are allowed for any $\chi_{\rm r}\in[-1/3,0)$.

With $n=4$ and $k=1$, the asymptotic solution~(\ref{H-asymp2})--(\ref{H3+beta-N}) near a {\it non-degenerate Killing horizon} for $\beta\in\mathbb{N}_0$, or equivalently $\chi_{\rm r}= -1/(1+2{\cal K})$, reduces to
\begin{align}
&H(x)\simeq \sum_{i=1}^{3+\beta}H_i\Delta^i,\label{H-asymp2-B}\\
&r(x)\simeq r_0+r_1\Delta+r_{2+\beta}\Delta^{2+\beta},\label{r-asymp2-B}\\
&\rho\simeq -\frac{2(2+\beta)(1+\beta)H_1r_{2+\beta}}{\kappa_4(1+\chi_{\rm r})r_0}\Delta^{1+\beta}, \label{rho-asymp2-20-B}
\end{align}
with $H_1\ne 0$, where 
\begin{align}
&H_1=\frac{1-\Lambda {r_0}^2}{r_0r_1},\qquad H_2=-\frac{1}{{r_0}^2},\label{H2-out-B}\\
&H_3=\frac{r_1(3- \Lambda {r_0}^2)}{3{r_0}^3}\quad (\mbox{for}~\beta\ge 1),\label{H3-out-B}\\
&H_{i+1}=-\frac{r_1}{r_0}H_{i}\quad (\mbox{with}~3\le i\le 1+\beta~\mbox{for}~\beta\ge 2),\label{H-recursion-B}
\end{align}
and
\begin{align}
&H_{3}=\frac{r_1(3-\Lambda {r_0}^2)}{3{r_0}^3}-\frac{2H_1r_{2}}{3r_0}\biggl(\frac{2\chi_{\rm t}}{1+\chi_{\rm r}} +2\biggl)~~\mbox{for}~~\beta=0,\\
&H_{4}=-\frac{{r_1}^2(3-\Lambda {r_0}^2)}{3{r_0}^4}-\frac{H_1r_{3}}{2r_0}\biggl(\frac{4\chi_{\rm t}}{1+\chi_{\rm r}} +3\biggl)~~\mbox{for}~~\beta=1,\\
&H_{3+\beta}=\frac{{r_1}^2}{{r_0}^2}H_{1+\beta} -\frac{2H_1r_{2+\beta}}{(3+\beta)r_0}\biggl\{\frac{2(1+\beta)\chi_{\rm t}}{1+\chi_{\rm r}} +(2+\beta)\biggl\}~~\mbox{for}~~\beta\ge 2, \label{H3+beta-N-B}
\end{align}
where $H_{1+\beta}$ on the right-hand side of Eq.~(\ref{H3+beta-N-B}) can be written in terms of $r_0$ and $r_1$ by Eqs.~(\ref{H3-out-B}) and (\ref{H-recursion-B}).
In Proposition~\ref{Prop:degenerate}, we showed that {\it degenerate Killing horizons} are allowed only for $(n-3)k\Lambda>0$.
For $\beta\in\mathbb{N}_0$ with $n=4$ and $k=1$, the asymptotic forms (\ref{H-asymp2-app})--(\ref{Crho-app2}) near a degenerate horizon reduce to
\begin{align}
&H(x)\simeq \sum_{i=2}^{4+2\beta}H_i\Delta^i,\label{H-asymp2-app-B}\\
&r(x)\simeq r_0+r_1\Delta+r_{2+2\beta}\Delta^{2+2\beta},\label{r-asymp2-app-B}\\
&\rho\simeq -\frac{2(2+2\beta)(1+2\beta)H_2r_{2+2\beta}}{\kappa_4(1+\chi_{\rm r})r_0}\Delta^{2+2\beta} \label{rho-asymp2-app-B}
\end{align}
with 
\begin{align}
\label{Crho-app2-B}
&H_2=-\frac{2\Lambda}{n-2}(\ne 0),\qquad r_0^2=\frac{(n-2)(n-3)k}{2\Lambda}.
\end{align}

Equations~(14)--(16) in Ref.~\cite{Bronnikov:2008ia} are written in the notation of the present paper as
\begin{align}
&H(x)\simeq H_{{\cal N}}\Delta^{{\cal N}}(1+o(\Delta^0)),\label{BZ-H}\\
&r(x)\simeq r_0+r_1\Delta+r_{2}\Delta^2+o(\Delta^2),\\
&\rho(x)\simeq \rho_{{\cal K}}\Delta^{{\cal K}}(1+o(\Delta^0)),
\end{align}
where ${\cal N}\in \mathbb{N}$ corresponds to $n$ in Ref.~\cite{Bronnikov:2008ia} and satisfies
\begin{align}
{\cal K}=-{\cal N}\frac{\chi_{\rm r}+1}{2\chi_{\rm r}}(={\cal N}(1+\beta))~~\Leftrightarrow~~\chi_{\rm r}=-\frac{{\cal N}}{{\cal N}+2{\cal K}}.\label{BZ}
\end{align}
Theorem~2 in Ref.~\cite{Bronnikov:2008ia} is a result in the case with non-vanishing $\rho_{\rm (vac)}$, $\rho_{\rm r(vac)}$, and $p_{\rm \perp(vac)}$, which corresponds to $\Lambda\ne 0$ in the present paper.
It asserts that (i) static solutions admit Killing horizons only for $\chi_{\rm r}=-{\cal N}/({\cal N}+2{\cal K})$ with ${\cal N}\le {\cal K}$ and then $\rho\propto \Delta^{{\cal K}}$ holds near a horizon.
Our results given by Eqs.~(\ref{H-asymp2-B})--(\ref{H3+beta-N-B}) for a non-degenerate horizon and by Eqs.~(\ref{H-asymp2-app-B})--(\ref{Crho-app2-B}) for a degenerate horizon correspond to ${\cal N}=1$ and ${\cal N}=2$, respectively.
Although analyticity of the metric is implicitly assumed in Theorem~2 in Ref.~\cite{Bronnikov:2008ia}, degenerate Killing horizons would also be allowed for $\chi_{\rm r}\ne -{\cal N}/({\cal N}+2{\cal K})$ with a non-analytic metric.
Besides, since $H''(x_{\rm h})\ne 0$ is required by Proposition~\ref{Prop:degenerate} in the present paper, a stronger condition $p_2\simeq \chi_{\rm t}\rho$ near the horizon prohibits ${\cal N}\ge 3$ in Eqs.~(\ref{BZ-H})--(\ref{BZ}), which corresponds to $n\ge 3$ in Eqs.~(14)--(16) in Ref.~\cite{Bronnikov:2008ia}.

\subsection{Applications to a perfect-fluid solution}

In this subsection, we apply our results to an exact solution as a demonstration.
In particular, we apply to the Semiz class-I solution with $k=1$, which is a solution with a perfect fluid obeying a linear equation of state $p=-(n-3)\rho/(n+1)$ without $\Lambda$~\cite{Semiz:2020lxj,Maeda:2022lsm}.
The solution is given by 
\begin{align}
\label{Semiz-I-twopara2}
\begin{aligned}
&\D s^2=-f\D {t}^2+h^{-2(n-4)/(n-3)}\left(f^{-1}\D y^2+y^2h^2\gamma_{ij}\D z^i\D z^j\right),\\
&\rho=-\frac{n+1}{n-3}p=\frac{(n^2-1)(n-2){\zeta}}{2(n-3)\kappa_n}f^{2/(n-3)}
\end{aligned} 
\end{align} 
with
\begin{align}
\begin{aligned}
&h(y)=1-\frac{{\zeta}}{y^{n-3}}\left(y^{n-3}-2M\right)^{(n-1)/(n-3)},\\
&f(y)=\frac{y^{n-3}-2M}{y^{n-3}h(y)}.
\end{aligned} 
\end{align} 
The solution is parametrized by $M$ and $\zeta$ and $\zeta=0$ gives the Schwarzschild-Tangherlini vacuum solution.
For $M>0$, a Killing horizon is located at $y=y_{\rm h}:=(2M)^{1/(n-3)}$.

The Buchdahl coordinate $x$ defined by Eq.~(\ref{metric-Buchdahl}) is given by 
\begin{align}
x=&\int\frac{\D y}{h(y)^{(n-4)/(n-3)}}.
\end{align} 
Then, we identify $H(x)\equiv f(y(x))$ and the areal radius $r$ is given by $r:=yh^{1/(n-3)}$.
The derivatives of the metric functions with respect to $x$ evaluated at the Killing horizon are given by 
\begin{align}
\begin{aligned}
H'|_{x=x_{\rm h}}=&\frac{n-3}{y_{\rm h}},\qquad H''|_{x=x_{\rm h}}=-\frac{(n-2)(n-3)}{y_{\rm h}^2},\qquad r'|_{x=x_{\rm h}}=1,\\
r''|_{x=x_{\rm h}}=&-\frac{2(n-1)\zeta}{n-3}y_{\rm h}^{n-4}(y^{n-3}-2M)^{-(n-5)/(n-3)}|_{y^{n-3}\to 2M},
\end{aligned} 
\end{align} 
where we used $\D/\D x=h^{(n-4)/(n-3)}\D/\D y$.

Now we reproduce the results in Ref.~\cite{Maeda:2022lsm}.
By Lemma~\ref{lm:finiteness-horizon} or Proposition~\ref{Prop:non-existence}, $y=y_{\rm h}$ is not a Killing horizon but a p.p. curvature singularity for $n\ge 6$ as $r''$ blows up.
In contrast, the solution admits a Killing horizon for $n=4$ ($p=-\rho/5$) and $n=5$ ($p=-\rho/3$), which is consistent with Proposition~\ref{Prop:asymp}.
Then, by Proposition~\ref{prop:matter-horizon}, there exists a non-vanishing matter field of the Hawking-Ellis type II on the horizon for $n=5$, while a matter field is absent on the horizon for $n=4$.
Since the values of $r$, $r'$, and $H'$ at the horizon do not depend on $\zeta$, two Semiz class-I solutions with different values of $\zeta$ can be attached at $y=y_{\rm h}$ in a $C^{1,1}$ regular manner for $n=4$ and $5$.
As a special case, one can attach the Semiz class-I solution to the Schwarzschild-Tangherlini vacuum solution with the same value of $M$.

\subsection{Generalization to Lovelock gravity}
\label{sec:Lovelock}

In this subsection, we extend Propositions~\ref{Prop:non-existence}, \ref{Pro:limit-horizon}, and~\ref{prop:matter-horizon} to Lovelock gravity with a maximally symmetric base manifold $K^{n-2}$.
Lovelock gravity is the most natural generalization of general relativity in arbitrary dimensions without torsion such that the field equations are of the second order~\cite{Lovelock:1971yv}, which gives rise to the ghost-free nature of the theory.

The action of Lovelock gravity is given by
\begin{align}
\label{action-L}
\begin{aligned}
&I_{\rm L}=\frac{1}{2\kappa_n}\int \D ^nx\sqrt{-g}\sum_{p=0}^{[n/2]}\alpha_{(p)}{\ma L}_{(p)}+I_{\rm matter},\\
&{\ma L}_{(p)}:=\frac{1}{2^p}\delta^{\mu_1\cdots \mu_p\nu_1\cdots \nu_p}_{\rho_1\cdots \rho_p\sigma_1\cdots \sigma_p}R_{\mu_1\nu_1}^{\phantom{\mu_1}\phantom{\nu_1}\rho_1\sigma_1}\cdots R_{\mu_p\nu_p}^{\phantom{\mu_p}\phantom{\nu_p}\rho_p\sigma_p},
\end{aligned}
\end{align}
where $I_{\rm matter}$ is the matter action to give an energy-momentum tensor $T_{\mu\nu}$ in the field equations.
The $\delta$ symbol in Eq.~(\ref{action-L}) is defined by 
\begin{align}
\delta^{\mu_1\cdots \mu_p}_{\rho_1\cdots \rho_p}:=&p!\delta^{\mu_1}_{[\rho_1}\cdots \delta^{\mu_p}_{\rho_p]},
\end{align}
which take values $0$ and $\pm 1$.
The constants $\alpha_{(p)}$ are coupling constants of the $p$th-order Lovelock terms and we have $\alpha_{(0)}=-2\Lambda$, where $\Lambda$ is a cosmological constant, and $\alpha_{(1)}=1$.
The first three Lovelock Lagrangian densities are given by 
\begin{align}
\begin{aligned}
{\ma L}_{(0)}=& 1,\qquad {\ma L}_{(1)}= R,\\
{\ma L}_{(2)}=& R^2-4R_{\mu\nu}R^{\mu\nu}+R_{\mu\nu\rho\sigma}R^{\mu\nu\rho\sigma}.
\end{aligned}
\end{align}
In fact, ${\ma L}_{(p)}$ is a dimensionally continued Euler density and therefore it does not contribute to the field equations if $p\ge [(n+1)/2]$ is satisfied.
As a consequence, the Lovelock action (\ref{action-L}) reduces to the Einstein-Hilbert action with $\Lambda$ for $n=3$ and $4$.

The gravitational field equations following from the action (\ref{action-L}) are given by 
\begin{align} 
{\ma G}_{\mu\nu}=&\kappa_n {T}_{\mu\nu}, \label{beqL}
\end{align} 
where ${\ma G}_{\mu\nu}$ is defined by
\begin{align} 
&{\ma G}_{\mu\nu} := \sum_{p=0}^{[n/2]}\alpha_{{(p)}}{G}^{(p)}_{\mu\nu},\label{generalG}\\
&{G}^{\mu(p)}_{~~\nu}:= -\frac{1}{2^{p+1}}\delta^{\mu\eta_1\cdots \eta_p\zeta_1\cdots \zeta_p}_{\nu\rho_1\cdots \rho_p\sigma_1\cdots \sigma_p}R_{\eta_1\zeta_1}^{\phantom{\eta_1}\phantom{\zeta_1}\rho_1\sigma_1}\cdots R_{\eta_p\zeta_p}^{\phantom{\eta_p}\phantom{\zeta_p}\rho_p\sigma_p}.
\end{align} 
Here ${G}^{(p)}_{\mu\nu}$ are the $p$th-order Lovelock tensors given from ${\ma L}_{(p)}$ and satisfy identities ${G}^{(p)}_{\mu\nu}\equiv 0$ for $p\ge [(n+1)/2]$.
By the $p$th-order Bianchi identities $\nabla^\nu{G}^{(p)}_{\mu\nu}\equiv 0$, the energy-momentum conservation equations $\nabla_\nu {T}^{\mu\nu}=0$ remain valid.

We assume that the base manifold $K^{n-2}$ is maximally symmetric for simplicity and use the result in Sec.~2.2 in Ref.~\cite{Maeda:2011ii} to write down the field equations.
It is because only a sub-class of the Einstein spaces is compatible with the Lovelock field equations (\ref{beqL}), and then the effect of the Weyl tensor of $K^{n-2}$ appears in the field equations in higher-order Lovelock gravity~\cite{Dotti:2005rc,Maeda:2010bu,Ohashi:2015xaa,Ray:2015ava}.
If $K^{n-2}$ is maximally symmetric, such terms do not appear and, by Eqs.~(2.18) and (2.21) in Ref.~\cite{Maeda:2011ii}, the energy-momentum tensor $T_{\mu\nu}$ compatible with the Lovelock field equations for a spacetime described by the metric (\ref{metric-Buchdahl}) is given in the type-I form~(\ref{T-diagonal}).

To simplify the equations, we will use
\begin{align}
&\Xi:=\frac{k-H{r'}^2 }{r^2},\label{def-Xi}\\
&{\tilde \alpha}_{(p)}:=\frac{(n-3)!\alpha_{(p)}}{(n-2p-1)!} \label{alphatil}
\end{align}
and
\begin{align}
&\Omega:=\sum_{p=0}^{[n/2]}p{\tilde \alpha}_{(p)}\Xi^{p-1},\label{def-Omega}\\
&X:=\sum_{p=0}^{[n/2]}(n-2p-1){\tilde \alpha}_{(p)}\Xi^{p}.\label{def-X}
\end{align}
First, we show that Propositions~\ref{Prop:non-existence} and \ref{Pro:limit-horizon} also hold in Lovelock gravity.

\begin{Prop}
\label{Prop:non-existence-L}
Propositions~\ref{Prop:non-existence} and \ref{Pro:limit-horizon} are valid in Lovelock gravity with a maximally symmetric base manifold $K^{n-2}$.
\end{Prop}
{\it Proof}. 
In the quasi-global coordinates (\ref{metric-Buchdahl}), ${\ma G}^\mu_{~\nu}$ is diagonal and then the field equations (\ref{beqL}) requires that the energy-momentum tensor $T^\mu_{~\nu}$ is also diagonal given by Eq.~(\ref{T-diagonal}).
Hence, we write down the $(t,t)$ and $(x,x)$ components of the field equations ${\ma G}^\mu_{~\nu}=\kappa_n T^\mu_{~\nu}$ as 
\begin{align}
\kappa_n\rho= & \frac{n-2}{2}\left[-r^{-1}(H'r'+2Hr'')\Omega + X\right],\label{Ltt}\\
\kappa_np_{\rm r} = & \frac{n-2}{2}\left(r^{-1}H'r'\Omega-X\right).\label{Lxx}
\end{align}
Adding Eqs.~(\ref{Ltt}) and (\ref{Lxx}), we obtain
\begin{align}
(n-2)r^{-1}Hr''\Omega=-\kappa_n(\rho+p_{\rm r}). \label{sum}
\end{align}
Equation~(\ref{sum}) shows that $\lim_{x\to x_{\rm h}}(p_{\rm r}+\rho)=0$ holds if $x=x_{\rm h}$ is a Killing horizon and hence Proposition~\ref{Pro:limit-horizon} is still valid. Because the conservation equation (\ref{conserve2}) holds in Lovelock gravity, we can follow the proof in Proposition~\ref{Prop:non-existence} and then Eq.~(\ref{ddr-proof-GR}) is modified to be 
\begin{align}
\Omega r''\simeq-\frac{\epsilon \kappa_n(1+\chi_{\rm r})r_0\Gamma_0}{n-2}|H|^{\beta},\label{ddr-proof-L}
\end{align}
which diverges for $\beta<0$ with $\chi_{\rm r}\ne 0,-1$, or equivalently, $\chi_{\rm r}\in(-\infty,-1)\cup (-1,-1/3)\cup(0,\infty)$.
Divergence of $r''\Omega$ means that at least either $r''$ or $r'$ blows up as $x\to x_{\rm h}$ and hence it is a curvature singularity by Lemma~\ref{lm:finiteness-horizon}.
\qed

Next, Proposition~\ref{prop:matter-horizon} is generalized as follows.
\begin{Prop}
\label{prop:matter-horizon-L}
Suppose that the metric (\ref{metric-Buchdahl-v}) with a maximally symmetric base manifold $K^{n-2}$ admits a Killing horizon $x=x_{\rm h}$ and solves the Lovelock field equations.
If $r''\Omega|_{x=x_{\rm h}}=0$ is satisfied, a matter field at $x=x_{\rm h}$ is in the Hawking-Ellis type-I form (\ref{T-typeI}) with $p_{\rm r}=-\rho$, which may be vanishing.
If $r''\Omega|_{x=x_{\rm h}}\ne 0$ is satisfied, there exists a non-vanishing matter field of the Hawking-Ellis type II at $x=x_{\rm h}$.
\end{Prop}
{\it Proof}. 
Non-zero components of the tensor (\ref{generalG}) for the metric (\ref{metric-Buchdahl-v}) are given by 
\begin{align}
{\cal G}^{vv} = & -(n-2)r^{-1}r''\Omega,\label{Gvv-L} \\
{\cal G}^{vx} = & {\cal G}^{xv} =\frac{n-2}{2}\left(r^{-1}H'r'\Omega-X\right),\label{Gvx-L} \\
{\cal G}^{xx} = &H{\cal G}^{vx},\qquad {\cal G}^{ij}=\kappa_n p_2r^{-2}\gamma^{ij},\label{Gij-L}
\end{align}
where $\kappa_np_2$ is given by 
\begin{align}
\kappa_n p_2=&\frac12\sum_{p=0}^{[n/2]}{\tilde\alpha}_{(p)}\Xi^{p-2}\biggl[-(n-2p-1)(n-2p-2)\Xi^{2}\nonumber\\
&+p\biggl\{H''+2(n-2p-1) \frac{H'r'+Hr''}{r}\biggl\} \Xi - p(p-1) \frac{H'r'(2Hr''+H'r')}{r^2}\biggl].\label{Lij}
\end{align}
Orthonormal component of the energy-momentum tensor $\kappa_nT^{(a)(b)}={\cal G}^{\mu\nu}E^{(a)}_{\mu}E^{(b)}_{\nu}$ with the orthonormal basis one-forms (\ref{basis0-H})--(\ref{basis2-H}) are computed to give
\begin{align}
\kappa_nT^{(0)(0)}=&\frac{n-2}{2}\biggl\{-\biggl[\left(1+\frac{H}{2}\right)^2\frac{r''}{r}+\frac{H'r'}{r}\biggl]\Omega+ X\biggl\},\label{T00-v-L}\\
\kappa_nT^{(0)(1)}=&\kappa_nT^{(1)(0)}=-\frac{n-2}{2}\left(1-\frac{H^2}{4}\right)\frac{r''}{r}\Omega,\\
\kappa_nT^{(1)(1)}=&\frac{n-2}{2}\biggl\{-\biggl[\left(1-\frac{H}{2}\right)^2\frac{r''}{r}-\frac{H'r'}{r}\biggl]\Omega- X\biggl\},\label{T11-v-L}\\
\kappa_nT^{(i)(j)}=& \kappa_np_2\delta^{(i)(j)}.\label{Tij-v-L}
\end{align}
Since $r''$ is finite at a Killing horizon $x=x_{\rm h}$ by Lemma~\ref{lm:finiteness-horizon}, $T^{(a)(b)}|_{x=x_{\rm h}}$ is in the Hawking-Ellis type-II form~(\ref{T-typeII}) with 
\begin{align}
&\rho_{\rm h}=\frac{n-2}{2\kappa_n}\left(-r^{-1}H'r'\Omega+ X\right)|_{x=x_{\rm h}},\qquad \nu=-\frac{n-2}{2\kappa_n}r^{-1}r''\Omega|_{x=x_{\rm h}},\label{T00-h-L}\\
&p_2|_{x=x_{\rm h}}=\sum_{p=0}^{[n/2]}\frac{{\tilde\alpha}_{(p)}}{2\kappa_n}\Xi^{p-2}\biggl[-(n-2p-1)(n-2p-2) \Xi^{2}\nonumber\\
&~~~~~~~~~~~~~~~~+p\biggl\{H''+2(n-2p-1) \frac{H'r'}{r}\biggl\} \Xi - p(p-1) \frac{{H'}^2{r'}^2}{r^2}\biggl]\biggl|_{x=x_{\rm h}},\label{Tij-h-L}
\end{align}
from which proposition follows.
Note that if $\nu=0$, namely $r''\Omega|_{x=x_{\rm h}}=0$, the energy-momentum tensor is in the type-I form (\ref{T-typeI}) with $p_{\rm r}=-\rho$.
\qed

\noindent
As Eqs.~(\ref{T00-v-L})--(\ref{T11-v-L}) give the following simple equation 
\begin{align}
&{\cal D}:=(T^{(0)(0)}+T^{(1)(1)})^2-4(T^{(0)(1)})^2=\frac{(n-2)^2H^2{r''}^2}{\kappa_n^2r^2}\Omega^2,\label{D-L}
\end{align}
one can use the criterion (\ref{D-criterion}) to identify the Hawking-Ellis type also in Lovelock gravity.

\section{Summary and concluding remarks}

In this paper, we have studied Killing horizons associated with a hypersurface-orthogonal Killing vector in $n(\ge 3)$-dimensional static spacetimes described by the metric (\ref{metric-Buchdahl}) or (\ref{metric-Buchdahl-v}) in general relativity with an $(n-2)$-dimensional Einstein base manifold and Lovelock gravity with a $(n-2)$-dimensional maximally symmetric base manifold. 
This work follows on from the four-dimensional studies ($n=4$) in Refs.~\cite{Pravda:2005uv,Bronnikov:2009ui} for general static spacetimes and in Refs.~\cite{Zaslavskii:2007pg,Bronnikov:2008ia,Bronnikov:2008by,Bronnikov:2011nb} for static spacetimes with spherically symmetry ($k=1$).
The only assumptions in the present paper are linear relations $p_{\rm r}\simeq \chi_{\rm r}\rho$ and $p_2\simeq \chi_{\rm t}\rho$ near the horizon between the radial pressure $p_{\rm r}$, tangential pressure $p_2$, and the energy density $\rho$ of the most general matter field compatible with the field equations.

In the present paper, we have defined the regularity of spacetime by a $C^{1,1}$ metric, which is sufficient to avoid curvature singularities and the divergence of matter fields in Lovelock gravity including general relativity as a special case.
In contrast, the analyticity ($C^\omega$) of the metric is assumed in Refs.~\cite{Bronnikov:2008ia,Bronnikov:2008by,Bronnikov:2009ui,Bronnikov:2011nb}.
We have generalized the previous studies for arbitrary $n(\ge 3)$ and $k(=1,0,-1)$ and the results have been stated in lemmas and propositions.
Among them, the following new results can be highlighted.
\begin{enumerate}
\item We have obtained a simple criterion to identify a p.p. curvature singularity. (See Lemma 3.) Remarkably, this criterion is useful for any spacetime described by the metric 
(\ref{metric-Buchdahl}) or (\ref{metric-Buchdahl-v}) in {\it any} gravitation theory.

\item For $\chi_{\rm r}<-1/3$ ($\chi_{\rm r}\ne -1$) or $\chi_{\rm r}>0$, static solutions do not admit Killing horizons in Lovelock gravity. (See Propositions~\ref{Prop:non-existence} and~\ref{Prop:non-existence-L}.)

\item For $\chi_{\rm r}=-1$, there exists a matter field of the Hawking-Ellis type~I on the Killing horizon in non-vacuum solutions in general relativity. (See Proposition~\ref{Prop:chi=-1}.)

\item For $\chi_{\rm r}\in[-1/3,0)$, non-vacuum solutions in general relativity admit non-degenerate Killing horizons and non-analytic extensions beyond the horizon are allowed.
In particular, such solutions can be attached to the Schwarzschild-Tangherlini-type vacuum solution at the horizon in at least a $C^{1,1}$ regular manner without a lightlike thin shell. (See Proposition~\ref{Prop:asymp}.) There exists a matter field on the horizon only for $\chi_{\rm r}=-1/3$, which is of the Hawking-Ellis type II. (See Proposition~\ref{Prop:matter-horizon}.)

\item Non-vacuum solutions in general relativity admit degenerate Killing horizons only for $(n-3)k\Lambda>0$ and then $H''(x_{\rm h})\ne 0$ holds. (See Proposition~\ref{Prop:degenerate}.)

\item If $p_{\rm r}=\chi_{\rm r}\rho$ and $p_2=\chi_{\rm t}\rho$ with $\chi_{\rm r}=-1$ or $\chi_{\rm r}\in[-1/3,0)$ are satisfied everywhere, not just near the horizon, the metric in non-vacuum solutions in general relativity is analytic on the horizon for $\chi_{\rm r}=-1$ and it can be $C^{\infty}$ there only for $\chi_{\rm r}= -1/(1+2N)$ with $N\in\mathbb{N}$.
(See Proposition~\ref{Prop:analytic-metric}.)
\end{enumerate}

\noindent
Some remarks about Lovelock gravity are in order.
In general relativity, the induced energy-momentum tensor of a lightlike thin shell localized on a $C^{0,1}$ junction null hypersurface can be computed by the Barrab\`{e}s-Israel junction conditions~\cite{Barrabes:1991ng,Poisson:2002nv}. However, such a method has not been established in general Lovelock gravity yet.
As stated in Propositions~\ref{Prop:non-existence-L} and~\ref{prop:matter-horizon-L}, Propositions~\ref{Prop:non-existence}, \ref{Pro:limit-horizon}, and~\ref{prop:matter-horizon} have been generalized in Lovelock gravity with a maximally symmetric base manifold $K^{n-2}$.
With a more general Einstein base manifold, the Weyl tensor ${}^{(n-2)}C_{ijkl}$ of the Einstein base manifold must satisfy certain algebraic conditions to be compatible with the Lovelock equations~\cite{Dotti:2005rc,Maeda:2010bu,Ray:2015ava,Ohashi:2015xaa}.
(See Eq.~(10) and the last equation in Sec.~4 in Ref.~\cite{Ray:2015ava}.)
We expect that Propositions~\ref{Prop:non-existence-L} and \ref{prop:matter-horizon-L} as well as Propositions~\ref{Prop:chi=-1}--\ref{Prop:analytic-metric} can be generalized in Lovelock gravity with such an Einstein base manifold, although the analysis is much more cumbersome.

In Proposition~\ref{Prop:asymp}, we have constructed asymptotic solutions near a non-degenerate Killing horizon for $\chi_{\rm r}\in[-1/3,0)$.
Then, the natural questions concerning those solutions are as follows.
\begin{itemize}

\item Such asymptotic solutions are certainly realized with a perfect fluid obeying a linear equation of state $p=\chi\rho$ for $\chi\in[-1/3,0)$. Can they be solutions with a fundamental field such as a (minimally or non-minimally coupled) scalar field or $p$-form field?

\item When asymptotic solutions are maximally extended, can such extended solutions describe an asymptotically flat static black hole satisfying, at least, the null energy condition? The answer is in fact negative in the Semiz class-I perfect-fluid solution obeying $p=-(n-3)\rho/(n+1)$ with $n=4$ and $5$ as it is asymptotically flat but admits an outer Killing horizon only with $\rho<0$~\cite{Maeda:2022lsm}.

\item As demonstrated for the Semiz class-I solution in Ref.~\cite{Maeda:2022lsm}, asymptotic solutions in the domain $x<x_{\rm h}$ can be attached to the exterior Schwarzschild-Tangherlini vacuum solution with positive mass and could even satisfy the null energy condition. The resulting black hole with a different interior cannot be distinguished from the Schwarzschild-Tangherlini black hole by observations. In addition, they share the same thermodynamical properties and dynamical stability of the exterior vacuum region. Then, what are the implications of such non-canonical configurations in the description of a black hole in quantum gravity?

\end{itemize}

Lastly, a natural and non-trivial future direction in this research is to extend our results to less symmetric spacetimes.
In particular, generalizations to the most general static spacetime and to stationary and axisymmetric spacetimes are worth challenging.
In fact, for a linear equation of state $p=\chi\rho$ with $\chi=-1/3$, an exact stationary and axisymmetric perfect-fluid solution with a Killing horizon is known in four~\cite{Wahlquist:1968zz,Senovilla1987} and higher dimensions~\cite{Hinoue:2014zta}.
We leave this task for future investigation.

\subsection*{Acknowledgements}
The authors are very grateful to Max-Planck-Institut f\"ur Gravitationsphysik (Albert-Einstein-Institut) for hospitality, where a large part of this work was carried out.
This work has been partially supported by the ANID FONDECYT grants 1201208, 1220862 and 1241835.

\appendix

\section{Corrections and complements to the analysis in Ref.~\cite{Bronnikov:2008by}}
\label{app:comple}

Thanks to Lemmas~\ref{lm:infinity}, \ref{lm:finiteness-horizon}, and \ref{lm:signature}, the analysis of $x=x_{\rm h}$ in the quasi-global coordinates (\ref{metric-Buchdahl}) is simple.
In this appendix, we provide several corrections and complements to the results in Sec.~IV in Ref.~\cite{Bronnikov:2008by} for $n=4$ with $k=1$, but the analysis below is valid for any $n(\ge 3)$ and $k$.

\begin{figure}[htbp]
\begin{center}
\includegraphics[width=0.7\linewidth]{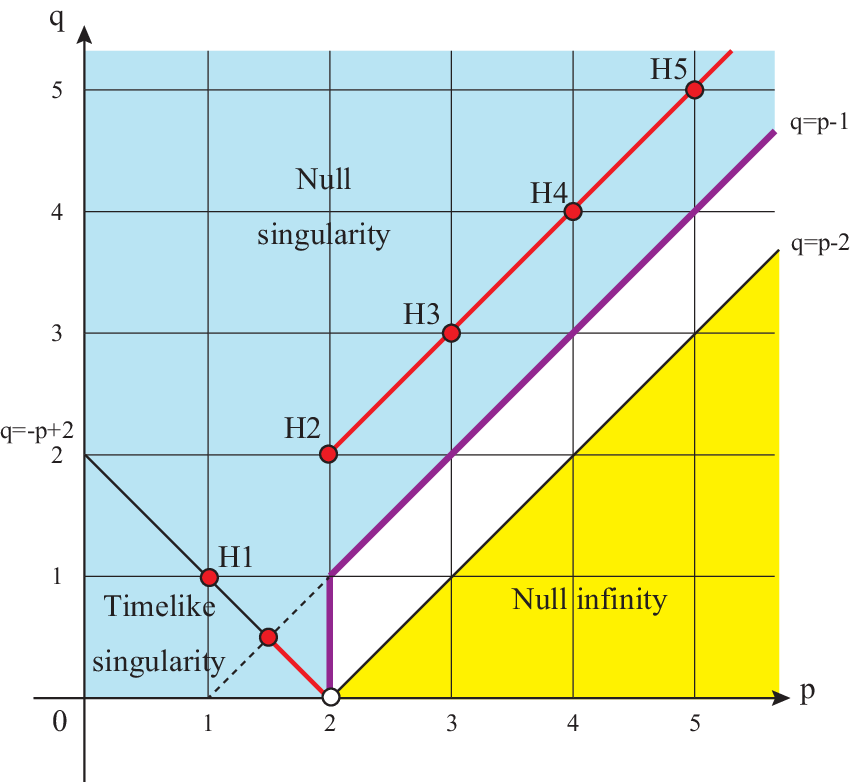}
\caption{\label{Fig-pq-plane} Properties of $r=r_{\rm h}$ depending on $p$ and $q$. It is a Killing horizon with at least a $C^{1,1}$ metric in the white region including the purple boundaries, while the other boundary $q=p-2$ with $p\ge 2$ belongs to null infinity. On the red point (H1) and the red segments, higher-order terms are necessary to check the regularity. This figure is compared with Fig.~1 in Ref.~\cite{Bronnikov:2008by}. }
\end{center}
\end{figure}

In Sec.~IV in Ref.~\cite{Bronnikov:2008by}, without specifying a theory, the authors studied the following metric in the areal coordinates
\begin{align}
\D s^2=&-e^{2\gamma(r)}\D t^2+e^{2\alpha(r)}\D r^2+r^2\gamma_{ij}(z)\D z^i\D z^j \label{Bronnikov}
\end{align}
and assumed
\begin{align}
e^{2\gamma}\simeq \gamma_0(r-r_{\rm h})^q,\qquad e^{2\alpha}\simeq \alpha_0(r-r_{\rm h})^{-p}
\end{align}
near $r=r_{\rm h}$ with a constant $p$ and positive constants $q$, $r_{\rm h}$, $\gamma_0$, and $\alpha_0$.
The metric (\ref{Bronnikov}) is written in the quasi-global coordinates (\ref{metric-Buchdahl}) with $H(x)\equiv e^{2\gamma(r(x))}$ and $r(x)$ that is the inverse of 
\begin{align}
x(r)=\int^r e^{\gamma(r)+\alpha(r)}\D r.
\end{align}
The finiteness of $x_{\rm h}:=x(r_{\rm h})$ requires $q>p-2$ and then we obtain
\begin{align}
x(r)\simeq x_{\rm h}+\frac{2\sqrt{\gamma_0\alpha_0}}{q-p+2}(r-r_{\rm h})^{(q-p+2)/2}
\end{align}
near $r=r_{\rm h}$, which gives
\begin{align}
&r(x)\simeq r_{\rm h}+\biggl(\frac{q-p+2}{2\sqrt{\gamma_0\alpha_0}}\biggl)^{2/(q-p+2)}(x-x_{\rm h})^{2/(q-p+2)},\label{r-asymp-past}\\
&H(x)\simeq \gamma_0(r-r_{\rm h})^q\simeq\gamma_0\biggl(\frac{q-p+2}{2\sqrt{\gamma_0\alpha_0}}\biggl)^{2q/(q-p+2)}(x-x_{\rm h})^{2q/(q-p+2)}.\label{H-asymp-past}
\end{align}
near $x=x_{\rm h}$.
For $q\le p-2$, $r=r_{\rm h}$ corresponds to $|x|\to \infty$ and hence it is a null infinity by Lemma~\ref{lm:infinity}.
This corresponds to the yellow region including the boundary $q=p-2$ in Fig.~\ref{Fig-pq-plane}, which is compared with Fig.~1 in Ref.~\cite{Bronnikov:2008by}.
In Ref.~\cite{Bronnikov:2008by}, despite that the authors realized that $r=r_{\rm h}$ corresponds to null infinity, it is referred to as a ``remote horizon'', which seems to be misleading.

It is important that the powers in Eqs.~(\ref{r-asymp-past}) and (\ref{H-asymp-past}) become one for $q=p$ and $q=-p+2$, respectively.
In those case, one has to consider higher-order terms to check the regularity of $r=r_{\rm h}$ such as
\begin{align}
&r(x)\simeq r_{\rm h}+\frac{1}{\sqrt{\gamma_0\alpha_0}}(x-x_{\rm h})+r_{1+\delta}(x-x_{\rm h})^{1+\delta} \label{higher-order-r}
\end{align}
or 
\begin{align}
&H(x)\simeq \frac{q\gamma_0}{\sqrt{\gamma_0\alpha_0}}(x-x_{\rm h})+H_{1+\eta}(x-x_{\rm h})^{1+\eta},
\end{align}
with positive constants $\delta$ and $\eta$.
However, this analysis is missing in Ref.~\cite{Bronnikov:2008by}.
In Sec.~IV in Ref.~\cite{Bronnikov:2008by}, the authors mentioned $C^2$ extensions at $r=r_{\rm h}$ in such a case without taking into account the contribution of higher-order terms.
In fact, for $0<\delta<1$ and $0<\eta<1$, $r''$ and $H''$ blow up as $r\to r_{\rm h}$, respectively, so that $r=r_{\rm h}$ is a curvature singularity by Lemma~\ref{lm:finiteness-horizon}.
In the present paper, it occurs for $-1<\beta<0$, or equivalently $-1<\chi_{\rm r}<-1/3$ as shown in Eqs.~(\ref{H-asymp1-app}) and (\ref{r-asymp1-app}).
The first and sixth columns in Table~I in Ref.~\cite{Bronnikov:2008by} state that $r=r_{\rm h}$ is regular for $q=p=N\in {\mathbb N}$, corresponding to the points ${\rm H}1, {\rm H}2,\cdots$ in Fig.~\ref{Fig-pq-plane}, and singular otherwise\footnote{Our $N$ is identical to $n$ used in Ref.~\cite{Bronnikov:2008by}.}.
However, this claim is not correct in general.
With $q=p=N$, we have Eq.~(\ref{higher-order-r}) and 
\begin{align}
H(x)\simeq \gamma_0\biggl(\frac{1}{\sqrt{\gamma_0\alpha_0}}\biggl)^{N}(x-x_{\rm h})^{N}+H_{N+\eta}(x-x_{\rm h})^{N+\eta}
\end{align}
near $x=x_{\rm h}$.
Thus, the correct statement is that $r=r_{\rm h}$ is regular and $C^2$ extendible for (i) $N=1$ with $\delta\ge 1$ and $\eta\ge 1$ or (ii) $N\ge 2$ with $\delta\ge 1$, and singular otherwise.

\begin{table*}[htb]
\begin{center}
\caption{Properties of $r=r_{\rm h}$ in Fig.~\ref{Fig-pq-plane}, which corrects Table~I in Ref.~\cite{Bronnikov:2008by}. In Nos. 1, 3, 4, and 6, properties depend on higher-order terms in the expansion.}
\label{table:corrected}
\begin{tabular}{|c|l|l|l|c|}\hline
No. & $p$, $q$ & Type by Eq.~(35) in Ref.~\cite{Bronnikov:2008by} & Present analysis \\ \hline
1 & $p=q=1$ & usual or naked & See the caption \\ \hline
2 & $1<p<3/2$, $q=-p+2$ & truly naked & nulll singularity \\ \hline
3 & $p=3/2$, $q=1/2$ & naked & See the caption \\ \hline
4 & $3/2<p<2$, $q=-p+2$ & usual & See the caption \\ \hline
5 & $p\ge 2$, $q>p$ & truly naked & null singularity \\ \hline
6 & $p\ge 2$, $q=p$ & usual or naked & See the caption \\ \hline
7 & $p\ge 2$, $p-1<q<p$ & truly naked & null singularity \\ \hline
8 & $p\ge 2$, $q=p-1$ & naked & regular \\ \hline
9 & $p\ge 2$, $p-2<q<p-1$ & usual & regular \\ \hline
10 & $p\ge 2$, $q\le p-2$ & usual & null infinity \\ \hline
\end{tabular}
\end{center}
\end{table*}

Now we assume $q>p-2$.
Then, $r''$ blows up as $r\to r_{\rm h}$ for $2/(q-p+2)<2$ with $2/(q-p+2)\ne 1$, which are identical to $q>p-1$ with $q\ne p$.
On the other hand, $H''$ blows up as $r\to r_{\rm h}$ for $2q/(q-p+2)<2$ and $2q/(q-p+2)\ne 1$, which are identical to $p<2$ with $q\ne -p+2$.
In those parameter regions, $r=r_{\rm h}$ is a curvature singularity by Lemma~\ref{lm:finiteness-horizon}, which is a light blue region in Fig.~\ref{Fig-pq-plane} not including the purple boundaries given by $q=p-1$ with $p\ge 2$ and $p=2$ with $0<q\le 1$, the red point $q=p=1$ (H1), and the red segments.
By Lemma~\ref{lm:infinity}, the singularity at $r=r_{\rm h}$ is non-null for $q<-p+2$ and null otherwise.
Because both $r''$ and $H''$ are finite, $r=r_{\rm h}$ is a Killing horizon in the parameter region given by $p-2 <q\le p-1$ and $p\ge 2$, which is a white region in Fig.~\ref{Fig-pq-plane} including the purple boundaries.

We summarize the results in Table~\ref{table:corrected} and it shows corrections to Table~I in Ref.~\cite{Bronnikov:2008by}, in which a null infinity is referred to as a ``remote horizon''.
For Nos.~1, 3, 4, and 6, a definitive conclusion cannot be obtained by the analysis adopted in that paper because it depends on higher-order terms in the expansion. 
Although Nos.~8 and 9 in Table~I in Ref.~\cite{Bronnikov:2008by} assert that $r=r_{\rm h}$ becomes singular for a certain range of $(p,q)$, it is actually regular with at least a $C^{1,1}$ metric. 
Truly naked horizons in Nos.~2, 5, and 7 are p.p. curvature singularities which are causally null.

\section{Solutions obeying $p_{\rm r}=\chi_{\rm r}\rho$ and $p_2=\chi_{\rm t}\rho$ with $\chi_{\rm r}=-1,0$}
\label{app:exact}
In this appendix, we study solutions to the Einstein equations for the metric (\ref{metric-Buchdahl}) with a matter field (\ref{T-diagonal}) obeying linear relations $p_{\rm r}=\chi_{\rm r}\rho$ and $p_2=\chi_{\rm t}\rho$.
Then, Eqs.~(\ref{EFE00})--(\ref{key}) are written as
\begin{align}
(n-2)r''H=&-\kappa_n(1+\chi_{\rm r})\rho r,\label{key-exact}\\
2\kappa_n\chi_{\rm r}\rho r^2=&(n-2)\left[rH'r'-(n-3)(k-H{r'}^2)\right]+2\Lambda r^2,\label{G-xx-exact}\\
2\kappa_n\chi_{\rm t}\rho r^2=&r^2H''+2(n-3)r(H'r'+Hr'')- (n-3)(n-4)(k-H{r'}^2)+2\Lambda r^2.\label{def-pt-exact}
\end{align}
The conservation equation~(\ref{conserve}) is integrated to give
\begin{align}
&\rho= \Theta_0r^{-(n-2)(\chi_{\rm r}-\chi_{\rm t})/\chi_{\rm r}} H^{1+\beta},\label{rho-exact}\\
&\beta:=-\frac{1+3\chi_{\rm r}}{2\chi_{\rm r}}\label{def-beta-exact}
\end{align}
for $\chi_{\rm r}\ne 0$ and 
\begin{align}
H=H_0r^{2(n-2)\chi_{\rm t}}\label{H-dust}
\end{align}
for $\chi_{\rm r}= 0$, where $\Theta_0$ and $H_0$ are integration constants. 
Therefore, the case with $\chi_{\rm r}=0$ must be treated separately.
This class of solutions with $\chi_{\rm r}=0$ has been obtained in Ref.~\cite{Biswas:2023omg} in general relativity and pure Lovelock gravity.

\noindent
{\bf Solution for $\chi_{\rm r}=-1$}

We present the most general static solution obeying $p_{\rm r}=-\rho$ and $p_2=\chi_{\rm t}\rho$, which generalizes the result in Ref.~\cite{Cho:2017nhx} for $n=4$, $k=1$, and $\Lambda=0$.
In the case of $\chi_{\rm r}=-1$, Eq.~(\ref{rho-exact}) reduces to
\begin{align}
\rho= \Theta_0r^{-(n-2)(1+\chi_{\rm t})}\label{rho+1}
\end{align}
and Eq.~(\ref{key-exact}) gives $r(x)=r_0+r_1 x$, where $r_0$ and $r_1$ are integration constant.
The solution is classified into two subclasses depending on whether $r_1$ is zero or not.

In the case of $r_1\ne 0$, we can set $r=x$ without loss of generality by a coordinate transformation $r_0+r_1 x\to x$.
Then, the general solution is given by 
\begin{align}
\label{chi=-1-sol1}
\begin{aligned}
&H(x)=k-\frac{2M}{x^{n-3}}+\frac{\eta}{x^{(n-2)\chi_{\rm t}+(n-4)}}-\frac{2\Lambda}{(n-1)(n-2)}x^2,\\
&r(x)=x,\qquad \rho=\frac{(n - 2)[(n-2)\chi_{\rm t}- 1]\eta}{2\kappa_n x^{(n - 2)(1+\chi_{\rm t})}}
\end{aligned}
\end{align}
for $\chi_{\rm t}\ne 1/(n-2)$ and 
\begin{align}
\label{chi=-1-sol1-special}
\begin{aligned}
&H(x)=k-\frac{2M}{x^{n-3}}+\frac{\eta\ln|x|}{x^{n-3}}-\frac{2\Lambda}{(n-1)(n-2)}x^2,\\
&r(x)=x,\qquad \rho=-\frac{(n - 2)\eta}{2\kappa_n x^{n-1}}
\end{aligned}
\end{align}
for $\chi_{\rm t}= 1/(n-2)$, where $M$ and $\eta$ are constants.
In the case of $r_1=0$, the general solution is the following direct-product solution
\begin{align}
\label{chi=-1-sol2}
\begin{aligned}
&H(x)=H_0+H_1x+H_2x^2,\\
&r(x)=r_0,\qquad \rho=\frac{(n-2)(n-3)k-2\Lambda r_0^2}{2\kappa_n r_0^2},
\end{aligned}
\end{align}
where $r_0$, $H_0$, and $H_1$ are arbitrary constants and $H_2$ is given by 
\begin{align}
&H_2=\frac{k(n-3)[(n - 2)\chi_{\rm t} + (n - 4)]- 2(\chi_{\rm t}+1)\Lambda r_0^2}{2r_0^2}.
\end{align}

\noindent
{\bf Solution for $\chi_{\rm r}=0$}

In the case of $\chi_{\rm r}=0$, substituting Eq.~(\ref{H-dust}) into Eq.~(\ref{G-xx-exact}), we obtain the master equation for $r(x)$:
\begin{align}
0=(n - 2)[2(n-2)\chi_{\rm t} + (n-3)]H_0r^{2(n - 2)\chi_{\rm t}}{r'}^2 + 2\Lambda r^2 - k(n - 2)(n - 3).\label{master-r-dust}
\end{align}
For $\chi_{\rm t}= -(n-3)/[2(n-2)]$, Eq.~(\ref{master-r-dust}) shows that $r$ is a constant given by 
\begin{align}
r=r_0:=\sqrt{\frac{k(n - 2)(n - 3)}{2\Lambda}},\label{r0-nariai}
\end{align}
which is real only for $k\Lambda>0$, and hence $H$ is also constant by Eq.~(\ref{H-dust}).
Then, Eq.~(\ref{EFE00}) gives $\rho\equiv 0$ and hence $p_2\equiv 0$.
Therefore, there is no non-vacuum solution for $\chi_{\rm t}= -(n-3)/[2(n-2)]$.

For $\chi_{\rm t}\ne -(n-3)/[2(n-2)]$, the master equation (\ref{master-r-dust}) is integrated to give
\begin{align}
&\pm(x-x_0)=\int\frac{\D r}{\sqrt{V(r)}},\label{eom}\\
&V(r):=\frac{k(n-2)(n-3)-2\Lambda r^2}{(n-2)[2(n-2)\chi_{\rm t}+(n-3)]H_0r^{2(n-2)\chi_{\rm t}}}.\label{V}
\end{align}
where $x_0$ is an integration constant.
The solution is real in domains where $V(r)>0$ holds and the energy density is given by
\begin{align}
\rho=&\frac{p_2}{\chi_{\rm t}}=\frac{k\chi_{\rm t} (n - 3)(n - 2)^2-2\Lambda[(n - 2)\chi_{\rm t}-1]r^2 }{\kappa_n[2(n - 2)\chi_{\rm t} +(n - 3)]r^2}.\label{rho-chir=0}
\end{align}
In the dust case $\chi_{\rm t}=0$, the energy density (\ref{rho-chir=0}) becomes constant such that
\begin{align}
\rho=&\frac{2\Lambda}{(n - 3)\kappa_n},
\end{align}
which shows that there is no non-vacuum solution with $\Lambda=0$ in the dust case.

For $\Lambda=0$, the energy density (\ref{rho-chir=0}) is given by 
\begin{align}
\rho=&\frac{p_2}{\chi_{\rm t}}=\frac{k\chi_{\rm t} (n - 3)(n - 2)^2}{\kappa_n[2(n - 2)\chi_{\rm t} +(n - 3)]r^2}
\end{align}
and the solution of the master equation (\ref{eom}) is 
\begin{align}
r(x)=&\biggl(\frac{k(n-3)[(n-2)\chi_{\rm t}+ 1]^2}{[2(n-2)\chi_{\rm t}+(n-3)]H_0}\biggl)^{1/[2\{(n-2)\chi_{\rm t}+1\}]} (x-x_0)^{1/[(n-2)\chi_{\rm t}+1]} 
\end{align}
for $\chi_{\rm t}\ne -1/(n-2)$ and 
\begin{align}
r(x)=\exp\biggl(\pm \sqrt{\frac{k(n-3)}{(n-5)H_0}}(x-x_0)\biggl)
\end{align}
for $\chi_{\rm t}=-1/(n-2)$.

\section{Asymptotic solutions near a non-degenerate Killing horizon for $\chi_{\rm r}\in[-1/3,0)$}
\label{app:coefficients}

In this appendix, we explain how to determine the asymptotic solutions in Proposition~\ref{Prop:asymp}.
We write Eqs.~(\ref{beq-asymp1})--(\ref{beq-asymp3}) as
\begin{align} 
&Hr''=-\frac{\kappa_n(p_{\rm r}+\rho)r}{n-2},\label{asymp1}\\
&H'rr'= (n-3)k-(n-3)H{r'}^2+\frac{2(\kappa_np_{\rm r}-\Lambda) r^2}{n-2}, \label{asymp2}\\
&2\kappa_np_2r^2=H''r^2+2(n-3)r(H' r'+Hr'')- (n-3)(n-4)(k-H{r'}^2)+2\Lambda r^2\label{asymp3}
\end{align}
and obtain power series solutions to Eqs.~(\ref{asymp1})--(\ref{asymp3}) up to the order of $|\Delta|^{1+\beta}$.

Equations~(\ref{r''-H})--(\ref{const1}) show that the asymptotic forms of the functions in the region $x>x_{\rm h}$ near a non-degenerate horizon $x=x_{\rm h}$ are given by 
\begin{align}
&H(x)\simeq \sum_{i=1}^{[3+\beta]}H_i\Delta^i+(\beta-[\beta])H_{3+\beta}\Delta^{3+\beta},\label{H-asymp1-app}\\
&r(x)\simeq r_0+r_1\Delta+r_{2+\beta}\Delta^{2+\beta},\label{r-asymp1-app}\\
&\rho\simeq \rho_{1+\beta} \Delta^{1+\beta} \label{rho-asymp1-app}
\end{align}
with $H_1\ne 0$ and 
\begin{align}
\rho_{1+\beta}=-\frac{(n-2)(2+\beta)(1+\beta)H_1r_{2+\beta}}{\kappa_n(1+\chi_{\rm r})r_0}.\label{Crho-app}
\end{align}

\noindent
{\bf For $\chi_{\rm r}=-1/(1+2N)$ with $N\in\mathbb{N}$}

First we consider the case of $\beta\in\mathbb{N}_0$, which is equivalent to $\chi_{\rm r}=-1/(1+2N)$ with $N\in\mathbb{N}$.
In this case, substituting Eqs.~(\ref{H-asymp1-app})--(\ref{Crho-app}) into Eq.~(\ref{asymp2}), we obtain
\begin{align}
&r_0r_1\sum_{i=0}^{1+\beta}(i+1)H_{i+1}\Delta^{i}\simeq (n-3)k-r_1^2\sum_{i=1}^{1+\beta}(n-3+i)H_i\Delta^i-\frac{2\Lambda}{n-2}(r_0+r_1\Delta)^2\label{master-H-int}
\end{align}
up to the order of $\Delta^{1+\beta}$.

In the case where at least $(n-3)k$ or $\Lambda$ is non-zero, Eq.~(\ref{master-H-int}) determines the coefficients in $H$ as
\begin{align}
&H_1=\frac{(n-2)(n-3)k-2\Lambda {r_0}^2}{(n-2)r_0r_1},\label{H1-out-app-N}\\
&H_2=\frac{-(n - 2)^2(n - 3)k + 2(n - 4)\Lambda {r_0}^2}{2(n-2){r_0}^2},\label{H2-out-app-N}\\
&H_3=\frac{(n-3)r_1\{(n-1)(n - 2)k - 2\Lambda {r_0}^2\}}{6{r_0}^3}\quad (\mbox{for}~\beta\ge 1),\label{H3-out-app-N}\\
&H_{i+1}=-\frac{(n-3+i)r_1}{(1+i)r_0}H_{i}\quad (\mbox{with}~3\le i\le 1+\beta~\mbox{for}~\beta\ge 2),\label{H-recursion-app-N}
\end{align}
while $H_{3+\beta}$ is undetermined.
Then, substituting those results into Eq.~(\ref{asymp3}), we obtain
\begin{align}
&2\kappa_n\chi_{\rm t}{r_0}^2\rho_{1+\beta} \Delta^{1+\beta} -2(n-3)(2+\beta)^2H_1r_0r_{2+\beta}\Delta^{1+\beta}\nonumber\\
&\simeq {r_0}^2\sum_{i=0}^{1+\beta}(i+2)(i+1)H_{i+2}\Delta^{i}+2r_0r_1\sum_{i=0}^{1+\beta}(n-3+i)(i+1)H_{i+1}\Delta^{i}\nonumber\\
&+{r_1}^2\sum_{i=1}^{1+\beta}\{(n-3)(n-4+2i)+i(i-1)\}H_i\Delta^i - (n-3)(n-4)k +2\Lambda (r_0+r_1\Delta)^2.\label{p2-key1}
\end{align}
We note that $\Delta^2$ terms contribute only if $\beta\ge 1$ and the terms with $\sum_{i=3}^{1+\beta}$ contribute only if $\beta\ge 2$.
For $\beta=0$, the terms of the order $\Delta^0$ are canceled by Eqs.~(\ref{H1-out-app-N}) and (\ref{H2-out-app-N}) and the terms of the order $\Delta^1$ give
\begin{align}
H_{3}=&\frac{(n-3)r_1\left\{(n-1)(n-2)k-2\Lambda {r_0}^2\right\}}{6{r_0}^3}-\frac{2H_1r_{2}}{3r_0}\biggl\{\frac{(n-2)\chi_{\rm t}}{1+\chi_{\rm r}} +2(n-3)\biggl\}.\label{H3-beta=0-N}
\end{align}
For $\beta=1$, the terms of the order $\Delta^0$ are canceled by Eqs.~(\ref{H1-out-app-N}) and (\ref{H2-out-app-N}), the terms of the order $\Delta^1$ are canceled by Eqs.~(\ref{H1-out-app-N})--(\ref{H3-out-app-N}), and the terms of the order $\Delta^2$ give
\begin{align}
H_{4}=&-\frac{n(n-3){r_1}^2\left\{(n-1)(n-2)k-2\Lambda {r_0}^2\right\}}{24{r_0}^4}-\frac{H_1r_{3}}{2r_0}\biggl\{\frac{2(n-2)\chi_{\rm t}}{1+\chi_{\rm r}} +3(n-3)\biggl\}.\label{H4-beta=1-N}
\end{align}
For $\beta\ge 2$, using Eqs.~(\ref{H1-out-app-N})--(\ref{H3-out-app-N}), we write Eq.~(\ref{p2-key1}) as
\begin{align}
&2\kappa_n\chi_{\rm t}{r_0}^2\rho_{1+\beta} \Delta^{1+\beta} -2(n-3)(2+\beta)^2H_1r_0r_{2+\beta}\Delta^{1+\beta}\nonumber\\
&\simeq{r_0}^2\sum_{i=3}^{\beta}(i+2)(i+1)H_{i+2}\Delta^{i}+{r_0}^2(3+\beta)(2+\beta)H_{3+\beta}\Delta^{1+\beta}\nonumber\\
&~~~+2r_0r_1\sum_{i=3}^{\beta}(n-3+i)(i+1)H_{i+1}\Delta^{i}+2r_0r_1(n-2+\beta)(2+\beta)H_{2+\beta}\Delta^{1+\beta}\nonumber\\
&~~~+{r_1}^2\sum_{i=3}^{\beta}\{(n-3)(n-4+2i)+i(i-1)\}H_i\Delta^i \nonumber\\
&~~~+{r_1}^2\{(n-3)(n-2+2\beta)+\beta(1+\beta)\}H_{1+\beta}\Delta^{1+\beta}.\label{p2-key1-2}
\end{align}
Since the terms with $\sum_{i=3}^{\beta}$ are canceled by Eq.~(\ref{H-recursion-app-N}), we obtain
\begin{align}
H_{3+\beta}=&\frac{{r_1}^2(n-1+\beta)(n-2+\beta)}{{r_0}^2(3+\beta)(2+\beta)}H_{1+\beta} \nonumber\\
&-\frac{2H_1r_{2+\beta}}{(3+\beta)r_0}\biggl\{\frac{(1+\beta)(n-2)\chi_{\rm t}}{1+\chi_{\rm r}} +(2+\beta)(n-3)\biggl\},\label{H3+beta-beta=2-N}
\end{align}
where $H_{1+\beta}$ on the right-hand side can be written in terms of $r_0$ and $r_1$ by Eqs.~(\ref{H3-out-app-N}) and (\ref{H-recursion-app-N}).
In this case, the parameters are $r_0$, $r_1$ (or $H_1$), and $r_{2+\beta}$.

For $(n-3)k=\Lambda=0$ with $H_1\ne 0$, Eq.~(\ref{master-H-int}) is satisfied for $r_1=0$.
Then, substituting the results into Eq.~(\ref{asymp3}), we obtain
\begin{align}
2\kappa_n\chi_{\rm t}r_0^2\rho_{1+\beta} \Delta^{1+\beta}\simeq& {r_0}^2\sum_{i=0}^{1+\beta}(i+2)(i+1)H_{i+2}\Delta^{i} +2(n-3)(2+\beta)^2r_0r_{2+\beta}H_1\Delta^{1+\beta}
\end{align}
up to the order of $\Delta^{1+\beta}$, which shows that
\begin{align}
\label{H3+beta-out}
\begin{aligned}
&H_2=H_3=\cdots=H_{1+\beta}=H_{2+\beta}=0,\\
&H_{3+\beta}=-\frac{2H_1r_{2+\beta}}{(3+\beta)r_0}\biggl\{(n-3)(2+\beta)+\frac{\chi_{\rm t}(n-2)(1+\beta)}{1+\chi_{\rm r}}\biggl\}.
\end{aligned}
\end{align}
In this case, the parameters are $r_0$, $r_{2+\beta}$, and $H_1(\ne 0)$.

\noindent
{\bf For $\chi_{\rm r}\ne -1/(1+2N)$ with $N\in\mathbb{N}$}

Next we consider a non-integer positive $\beta$, which is equivalent to $\chi_{\rm r}\ne -1/(1+2N)$ with $N\in\mathbb{N}$.
We study the domains $x\ge x_{\rm h}$ and $x\le x_{\rm h}$ separately.
In the domain $x\ge x_{\rm h}$, substituting Eqs.~(\ref{H-asymp1-app})--(\ref{Crho-app}) into Eq.~(\ref{asymp2}), we obtain
\begin{align}
r_0r_1\sum_{i=0}^{[1+\beta]}(i+1)H_{i+1}\Delta^{i}\simeq& (n-3)k-r_1^2\sum_{i=1}^{[1+\beta]}(n-3+i)H_i\Delta^i \nonumber\\
&-\frac{2\Lambda}{n-2}(r_0^2+2r_0r_1\Delta+{r_1}^2\Delta^2)\label{master-H}
\end{align}
up to the order of $\Delta^{1+\beta}$.

In the case where at least $(n-3)k$ or $\Lambda$ is non-zero, Eq.~(\ref{master-H}) determines the coefficients in $H$ as
\begin{align}
&H_1=\frac{(n-2)(n-3)k-2\Lambda {r_0}^2}{(n-2)r_0r_1},\label{H1-out-app}\\
&H_2=\frac{-(n - 2)^2(n - 3)k + 2(n - 4)\Lambda {r_0}^2}{2(n-2){r_0}^2},\label{H2-out-app}\\
&H_3=\frac{(n-3)r_1\{(n-1)(n - 2)k - 2\Lambda {r_0}^2\}}{6{r_0}^3}\quad (\mbox{for}~\beta> 1),\label{H3-out-app}\\
&H_{i+1}=-\frac{(n-3+i)r_1}{(1+i)r_0}H_{i}\quad (\mbox{for}~3\le i\le [1+\beta]~\mbox{if}~\beta> 2),\label{H-recursion-app}
\end{align}
while $H_{[3+\beta]}$ and $H_{3+\beta}$ are undetermined.
Then, substituting those results into Eq.~(\ref{asymp3}), we obtain
\begin{align}
&2\kappa_n\chi_{\rm t}{r_0}^2\rho_{1+\beta} \Delta^{1+\beta} -2(n-3)(2+\beta)^2H_1r_0r_{2+\beta}\Delta^{1+\beta}-(3+\beta)(2+\beta){r_0}^2H_{3+\beta}\Delta^{1+\beta}\nonumber\\
&\simeq 6{r_0}^2H_{3}\Delta+4(n-2)r_0r_1H_{2}\Delta+(n-3)(n-2){r_1}^2H_1\Delta+4\Lambda r_0r_1\Delta \nonumber\\
&~~~+12{r_0}^2H_{4}\Delta^2+6(n-1)r_0r_1H_{3}\Delta^{2}+(n-1)(n-2){r_1}^2H_2\Delta^2+2\Lambda {r_1}^2\Delta^2\nonumber\\
&~~~+{r_0}^2\sum_{i=3}^{[1+\beta]}(i+2)(i+1)H_{i+2}\Delta^{i}+2r_0r_1\sum_{i=3}^{[1+\beta]}(n-3+i)(i+1)H_{i+1}\Delta^{i}\nonumber\\
&~~~+{r_1}^2\sum_{i=3}^{[1+\beta]}\{(n-3)(n-4+2i)+i(i-1)\}H_i\Delta^i \label{p2-key2}
\end{align}
up to the order of $\Delta^{1+\beta}$.
We note that $\Delta^2$ terms contribute only if $\beta> 1$ and the terms with $\sum_{i=3}^{[1+\beta]}$ contribute only if $\beta> 2$.
From the terms of the order $\Delta^{1+\beta}$, we obtain
\begin{align}
H_{3+\beta}=&-\frac{2H_1r_{2+\beta}}{(3+\beta)r_0}\biggl\{(n-3)(2+\beta)+\frac{(n-2)(1+\beta)\chi_{\rm t}}{1+\chi_{\rm r}}\biggl\}.\label{H3+beta-correct}
\end{align}
The expression of $H_{[3+\beta]}$ is obtained as follows.
For $0<\beta<1$, the terms of the order $\Delta^1$ with Eqs.~(\ref{H1-out-app}) and (\ref{H2-out-app}) give
\begin{align}
&H_3(=H_{[3+\beta]})=\frac{(n-3)r_1\{(n-1)(n - 2)k - 2\Lambda {r_0}^2\}}{6{r_0}^3}\quad (\mbox{for}~0<\beta<1).\label{H3-0<beta<1}
\end{align}
Then, for $1<\beta<2$, the terms of the order $\Delta^2$ with Eqs.~(\ref{H1-out-app})--(\ref{H3-out-app}) give
\begin{align}
H_4(=H_{[3+\beta]})=&-\frac{n(n-3){r_1}^2\{(n-1)(n - 2)k - 2\Lambda {r_0}^2\}}{24{r_0}^4} \nonumber\\
=&-\frac{nr_1}{4r_0}H_3\quad (\mbox{for}~1<\beta<2).\label{H4-1<beta<2}
\end{align}
For $\beta>2$, the terms of the order $\Delta^{[1+\beta]}$ give
\begin{align}
H_{[3+\beta]}=&-\frac{2r_1[n-2+\beta]}{r_0[3+\beta]}H_{[2+\beta]} \nonumber\\
&-\frac{{r_1}^2\{(n-3)(n-4+2[1+\beta])+[1+\beta][\beta]\}}{{r_0}^2[3+\beta][2+\beta]}H_{[1+\beta]}.\label{H[3+beta]-g}
\end{align}
In this case, the parameters are $r_0$, $r_1$ (or $H_1$), and $r_{2+\beta}$.

For $(n-3)k=\Lambda=0$ with $H_1\ne 0$, Eq.~(\ref{master-H}) is satisfied for $r_1=0$.
Then, substituting those results into Eq.~(\ref{asymp3}), we obtain
\begin{align}
2\kappa_n\chi_{\rm t}r_0^2\rho_{1+\beta}\Delta^{1+\beta}\simeq& {r_0}^2\sum_{i=0}^{[1+\beta]}(i+2)(i+1)H_{i+2}\Delta^{i} +(2+\beta)(1+\beta){r_0}^2H_{2+\beta}\Delta^{\beta} \nonumber\\
&+(3+\beta)(2+\beta){r_0}^2H_{3+\beta}\Delta^{1+\beta} +2(n-3)(2+\beta)^2r_0r_{2+\beta}H_1\Delta^{1+\beta}
\end{align}
up to the order of $\Delta^{1+\beta}$, which shows
\begin{align}
&H_2=H_3=\cdots=H_{[2+\beta]}=H_{[3+\beta]}=0
\end{align}
and $H_{3+\beta}$ given by Eq.~(\ref{H3+beta-out}).
In this case, the parameters are $r_0$, $r_{2+\beta}$, and $H_1(\ne 0)$.

In the domain $x\le x_{\rm h}$, substituting the expansions
\begin{align}
&H(x)\simeq \sum_{i=1}^{[3+\beta]}H_i\Delta^i +{\bar H}_{3+\beta}(-\Delta)^{3+\beta},\label{expansion-H1-in-app}\\
&r(x)\simeq r_0+r_1\Delta+{\bar r}_{2+\beta}(-\Delta)^{2+\beta}\label{expansion-r1-in-app}
\end{align}
with $H_1\ne 0$ into Eq.~(\ref{asymp1}), we obtain 
\begin{align}
&\rho\simeq {\bar \rho}_{1+\beta}(-\Delta)^{1+\beta},\label{expansion-rho1-in-app}\\
&{\bar \rho}_{1+\beta}=\frac{(n-2)(2+\beta)(1+\beta)H_1{\bar r}_{2+\beta}}{\kappa_n(1+\chi_{\rm r})r_0}\label{Crho-in-app}
\end{align}
Then, substituting Eqs.~(\ref{expansion-H1-in-app})--(\ref{Crho-in-app}) into Eq.~(\ref{asymp2}), we obtain
\begin{align}
r_0r_1\sum_{i=0}^{[1+\beta]}(i+1)H_{i+1}\Delta^{i} \simeq &(n-3)k-{r_1}^2\sum_{i=1}^{[1+\beta]}(n-3+i)H_i\Delta^i\nonumber\\
&-\frac{2\Lambda}{n-2}({r_0}^2+2r_0r_1\Delta+{r_1}^2\Delta^2)\label{master-H-in}
\end{align}
up to the order of $(-\Delta)^{1+\beta}$.

In the case where at least $(n-3)k$ or $\Lambda$ is non-zero, Eq.~(\ref{master-H-in}) determines the coefficients in $H$ as Eqs.~(\ref{H1-out-app})--(\ref{H-recursion-app}).
Then, substituting those results into Eq.~(\ref{asymp3}), we obtain
\begin{align}
&2\kappa_n\chi_{\rm t}{r_0}^2{\bar \rho}_{1+\beta}(-\Delta)^{1+\beta}+2(n-3)(2+\beta)^2H_1r_0{\bar r}_{2+\beta}(-\Delta)^{1+\beta} \nonumber\\
&-(3+\beta)(2+\beta){r_0}^2{\bar H}_{3+\beta}(-\Delta)^{1+\beta}\nonumber\\
\simeq&6{r_0}^2H_{3}\Delta+4(n-2)r_0r_1H_{2}\Delta+(n-3)(n-2){r_1}^2H_1\Delta+4\Lambda r_0r_1\Delta \nonumber\\
&~~~+12{r_0}^2H_{4}\Delta^2+6(n-1)r_0r_1H_{3}\Delta^{2}+(n-1)(n-2){r_1}^2H_2\Delta^2+2\Lambda {r_1}^2\Delta^2\nonumber\\
&~~~+{r_0}^2\sum_{i=3}^{[1+\beta]}(i+2)(i+1)H_{i+2}\Delta^{i}+2r_0r_1\sum_{i=3}^{[1+\beta]}(n-3+i)(i+1)H_{i+1}\Delta^{i}\nonumber\\
&~~~+{r_1}^2\sum_{i=3}^{[1+\beta]}\{(n-3)(n-4+2i)+i(i-1)\}H_i\Delta^i \label{p2-key3}
\end{align}
up to the order of $(-\Delta)^{1+\beta}$.
From the terms of the order $\Delta^{1+\beta}$, we obtain
\begin{align}
{\bar H}_{3+\beta}=&\frac{2H_1{\bar r}_{2+\beta}}{(3+\beta)r_0}\biggl\{(n-3)(2+\beta)+\frac{(n-2)(1+\beta)\chi_{\rm t}}{1+\chi_{\rm r}}\biggl\}.\label{H3+beta-correct2}
\end{align}
Since the right-hand sides of Eqs.~(\ref{p2-key2}) and (\ref{p2-key3}) are identical, Eqs.~(\ref{H3-0<beta<1})--(\ref{H[3+beta]-g}) are valid as the expressions of $H_{[3+\beta]}$ for $0<\beta<1$, for $1<\beta<2$, and for $\beta>2$, respectively.
In this case, the parameters are $r_0$, $r_1$ (or $H_1$), and ${\bar r}_{2+\beta}$.

For $(n-3)k=\Lambda=0$ with $H_1\ne 0$, Eq.~(\ref{master-H-in}) is satisfied for $r_1=0$.
Then, substituting those results into Eq.~(\ref{asymp3}), we obtain
\begin{align}
2\kappa_n\chi_{\rm t}{r_0}^2{\bar \rho}_{1+\beta}(-\Delta)^{1+\beta}\simeq &{r_0}^2\sum_{i=0}^{[1+\beta]}(i+2)(i+1)H_{i+2}\Delta^{i}+(3+\beta)(2+\beta){r_0}^2{\bar H}_{3+\beta}(-\Delta)^{1+\beta} \nonumber\\
&-2(n-3)(2+\beta)^2r_0 {\bar r}_{2+\beta}H_1(-\Delta)^{1+\beta}
\end{align}
up to the order of $(-\Delta)^{1+\beta}$.
For $\beta\notin\mathbb{N}$, the above equation shows
\begin{align}
&H_2=H_3=\cdots=H_{[2+\beta]}=H_{[3+\beta]}=0
\end{align}
and ${\bar H}_{3+\beta}$ given by Eq.~(\ref{H3+beta-correct2}).
In this case, the parameters are $r_0$, ${\bar r}_{2+\beta}$, and $H_1(\ne 0)$.

\section{A refined statement of Proposition~6}
\label{addendum}

\setcounter{Prop}{5}

This appendix is based on the addendum~\cite{Addendum-CQG} to the published version.
In Proposition~6, we assumed $r_0>0$ but $r_0\ne 0$ is sufficient.
More importantly, we implicitly assumed that $\chi_{\rm r}$ and $\chi_{\rm t}$ are the same on both sides of the horizon.
However, as $\beta\ge 0$ holds for $\chi_{\rm r}\in[-1/3,0)$, the $C^{1,1}$ regularity at $x=x_{\rm h}$ is guaranteed if the values of $r_0$ and $H_1$ are the same on both sides of the horizon.
Actually, in the case where at least $(n-3)k$ or $\Lambda$ is non-zero, the constraint~(\ref{H1-out}) between $r_0$, $r_1$, and $H_1$ is independent from $\chi_{\rm r}$ and $\chi_{\rm t}$.
In the case of $(n-3)k=\Lambda=0$, $r_0$ and $H_1$ are parameters.
Therefore, two solutions even with different values of $\chi_{\rm r}$ and $\chi_{\rm t}$ can be attached regularly at the horizon $x=x_{\rm h}$ regularly if $r_0$ and $H_1$ are the same on both sides of the horizon. 
Here we can set $H_1$ any non-zero value by scaling transformations of $t$ and $x$ and reparameterizations of other coefficients in the asymptotic solution.
We summarize this result in the following refined statement of Proposition~6.
\begin{Prop}
\label{Prop:asymp-new}
Consider spacetimes $({\cal M}_n^+,g_{\mu\nu}^+)$ and $({\cal M}_n^-,g_{\mu\nu}^-)$ as solutions to the Einstein equations (\ref{EFE-0}) defined in the domains $x>x_{\rm h}$ and $x<x_{\rm h}$, respectively, in the quasi-global coordinates (\ref{metric-Buchdahl}).
Suppose that (i) the components of the energy-momentum tensor (\ref{T-diagonal}) on $({\cal M}_n^\pm,g_{\mu\nu}^\pm)$ obey $p_{\rm r}\simeq \chi_{\rm r}^\pm\rho$ and $p_2\simeq \chi_{\rm t}^\pm\rho$ as $x\to x_{\rm h}$ with constants $\chi_{\rm r}^\pm\in[-1/3,0)$ and $\chi_{\rm t}^\pm$, and (ii) the solutions admit a non-degenerate Killing horizon $x=x_{\rm h}$.
Then, the solutions sharing the same value of $r(x_{\rm h})(\ne 0)$ can be attached at $x=x_{\rm h}$ in a regular manner with at least a $C^{1,1}$ metric in the single-null coordinates (\ref{metric-Buchdahl-v}).
\end{Prop}


Lastly, we present a concrete example of the claim in Proposition~\ref{Prop:asymp-new} in the four-dimensional ($n=4$) spherically symmetric case ($k=1$) with a perfect fluid obeying $p=\chi\rho$ using the Semiz class-I solution~\cite{Semiz:2020lxj} for $\chi=-1/5$ ($\beta=1$) and the Whittaker solution~\cite{Whittaker} for $\chi=-1/3$ ($\beta=0$).
In Refs.~\cite{Maeda:2022lsm,Maeda:2024tpl}, the Semiz class-I solution~\cite{Semiz:2020lxj} is presented in the comoving quasi-global coordinates (\ref{metric-Buchdahl}) as
\begin{align}
\begin{aligned}
&H(x)=\frac{x-2m}{x-{\zeta}\left(x-2m\right)^{3}},\\
&r(x)=x-{\zeta}\left(x-2m\right)^3,\\
&\rho=-5p=\frac{15{\zeta}}{\kappa}H^2.
\end{aligned}
\end{align} 
The solution is parametrized by $m$ and $\zeta$ and admits a single non-degenerate Killing horizon at $x=x_{\rm h}(=2m)$ for $m\ne 0$.
We compute
\begin{align}
\label{values-Semiz}
\begin{aligned}
&H(x_{\rm h})=0,\qquad r(x_{\rm h})=r_0,\\
&H'(x_{\rm h})={r_0}^{-1},\qquad r'(x_{\rm h})=1,\\
&H''(x_{\rm h})=-2{r_0}^{-2},\qquad r''(x_{\rm h})=0,
\end{aligned} 
\end{align} 
where $r_0:=r(x_{\rm h})=2m$~\cite{Maeda:2024tpl}.

On the other hand, the expression of the Whittaker solution in the comoving quasi-global coordinates (\ref{metric-Buchdahl}) depends on the parameters of the solution $M$ and $\alpha(\ne 0)$ and we focus on the case with $\alpha>0$ or $-1/(4M^2)<\alpha<0$~\cite{Maeda:2024tpl}.
For $\alpha>0$, the solution is written as
\begin{align}
\label{Hr+}
\begin{aligned}
&H(x)=\frac{1}{\omega}\biggl(1-\frac{2\sqrt{\alpha}M}{\tan(\sqrt{\alpha\omega}x)}\biggl),\\
&r(x)=\frac{1}{\sqrt{\alpha}}\sin(\sqrt{\alpha\omega}x),\\
&\rho=-3p=\frac{3\alpha\omega}{\kappa}H,
\end{aligned} 
\end{align} 
where we have introduced a gauge constant $\omega(>0)$ for later use.
For $-1/(4M^2)<\alpha<0$, the solution is written as
\begin{align}
\label{Hr-}
\begin{aligned}
&H(x)=\frac{1}{\omega}\biggl(1-\frac{2\sqrt{|\alpha|}M}{\tanh(\sqrt{|\alpha|\omega}x)}\biggl),\\
&r(x)=\frac{1}{\sqrt{|\alpha|}}\sinh(\sqrt{|\alpha|\omega}x),\\
&\rho=-3p=\frac{3\alpha\omega}{\kappa}H.
\end{aligned} 
\end{align} 
The solution with $M\ne 0$ admits a non-degenerate Killing horizon at $x=x_{\rm h}$, where
\begin{align}
&x_{\rm h}=\left\{
\begin{array}{ll}
\arctan(2\sqrt{\alpha}M)/\sqrt{\alpha\omega} & [\mbox{for}~\alpha>0]\\
{\rm arctanh}(2\sqrt{|\alpha|}M)/\sqrt{|\alpha|\omega} & [\mbox{for}~-1/(4M^2)<\alpha<0]
\end{array}
\right..\label{def-alpha}
\end{align} 
As in Ref.~\cite{Maeda:2024tpl}, we choose the gauge constant $\omega$ such that $r'(x_{\rm h})=1$, namely
\begin{align}
\omega=\frac{1}{1-\alpha r_0^2}=1+4\alpha M^2(>0),
\end{align} 
where 
\begin{align}
r_0:=r(x_{\rm h})=\frac{2M}{\sqrt{1+4\alpha M^2}}.\label{rh-W}
\end{align}
Then, we compute
\begin{align}
\label{values-W}
\begin{aligned}
&H(x_{\rm h})=0,\qquad r(x_{\rm h})=r_0,\\
&H'(x_{\rm h})={r_0}^{-1},\qquad r'(x_{\rm h})=1,\\
&H''(x_{\rm h})=-2{r_0}^{-2},\qquad r''(x_{\rm h})=-4\alpha M^2{r_0}^{-1}.
\end{aligned} 
\end{align} 

Now consider regular attachments of the Semiz class-I and Whittaker solutions at the Killing horizon. 
We can set $x_{\rm h}=0$ in both solutions by a shift transformation $x\to x+x_{\rm h}$, under which Eqs.~(\ref{values-Semiz}) and (\ref{values-W}) are unchanged.
Continuity of $r(x)$ at $x=x_{\rm h}(=0)$ requires the following condition
\begin{align}
m= \frac{M}{\sqrt{1+4\alpha M^2}},\label{condition-attach}
\end{align} 
under which the difference between Eqs.~(\ref{values-Semiz}) and (\ref{values-W}) is only $r''(x_{\rm h})$.
Therefore, under the condition (\ref{condition-attach}), the Semiz class-I solution and the Whittaker solution can be attached at the Killing horizon $x=x_{\rm h}(=0)$ in a regular manner with a $C^{1,1}$ metric in the single-null coordinates (\ref{metric-Buchdahl-v}).


\end{document}